Preprinted manuscript

# A potential network structure of symbiotic bacteria involved in carbon and nitrogen metabolism of wood-utilizing insect larvae

Old title: Symbiotic bacterial network structure involved in carbon and nitrogen metabolism of wood-utilizing insect larvae

Hirokuni Miyamoto*[1,2,3,4], Futo Asano[1], Koutarou Ishizawa[5], Wataru Suda[2], Hisashi Miyamoto[6], Naoko Tsuji[3], Makiko Matsuura[1,3], Arisa Tsuboi[3,4,7], Chitose Ishii[1,2,3], Teruno Nakaguma[1,3,4], Chie Shindo[2], Tamotsu Kato[2], Atsushi Kurotani[7], Hideaki Shima[7], Shigeharu Moriya[1,7], Masahira Hattori[2,8], Hiroaki Kodama[1], Hiroshi Ohno[2], Jun Kikuchi*[7]

*Affiliations:*
*1. Graduate School of Horticulture, Chiba University, Matsudo, Chiba 271-8501, Japan*
*2. RIKEN IMS, Yokohama, Kanagawa 230-0045, Japan*
*3. Sermas Co., Ltd., Ichikawa, Chiba 272-0033, Japan*
*4. Japan Eco-science (Nikkan Kagaku) Co., Ltd., Chiba, Chiba 260-0034, Japan*
*5. Elementary School of Batou, Nakagawa, Tochigi 324-0613, Japan*
*6. Miroku Co., Ltd., Kitsuki, Oita 873-0021, Japan*
*7. RIKEN CSRS, Yokohamai, Kanagawa 230-0045, Japan*
*8. School of Advanced Science and Engineering, Waseda University, Tokyo169-8555, Japan*

* Co-corresponding authors
Correspondence: Hirokuni Miyamoto Ph.D., Chiba University, RIKEN, Sermas Co., Ltd., and Japan Eco-science Co., Ltd.
Tel: +81-43-290-3947, Fax: +81-43-290-3947
E-mail: hirokuni.miyamoto@riken.jp, h-miyamoto@faculty.chiba-u.jp

Correspondence: Jun Kikuchi Ph.D., RIKEN
Tel: +81-43-290-3942, Fax: +81-43-290-3942
E-mail: jun.kikuchi@riken.jp

Abstract

Effective biological utilization of wood biomass is necessary worldwide. Since several insect larvae can use wood biomass as a nutrient source, studies on their digestive mechanisms are expected to reveal a novel rule underlying wood biomass processing. Here, the relationships of inhabitant bacteria involved in carbon and nitrogen metabolism in the intestine of beetle larvae, an insect model, are investigated. Bacterial analysis of larval feces showed enrichment of the phyla Chroloflexi, Gemmatimonadetes, Planctomycetes, the genera *Bradyrhizobium*, *Chonella*, *Corallococcus*, *Gemmata*, *Hyphomicrobium*, *Lutibacterium*, *Paenibacillus*, and *Rhodoplanes*, bacteria candidates for plant growth promotion, nitrogen cycle modulation, and/or environmental protection. The abundances of these bacteria were not necessarily positively correlated with their abundances in the habitat, indicating that they are selectively enriched in the feces of larvae. Therefore, association analysis predicted that common fecal bacteria might affect carbon and nitrogen metabolism. Based on the predicted hypotheses, structural equation modeling (SEM) statistically estimated that inhabitant bacterial groups involved in carbon and nitrogen metabolism were partly composed of candidate environmentally beneficial bacteria enriched in the feces. Still, the selected common bacteria, *i.e.,* the phyla Acidobacteria, Armatimonadetes, Bacteroidetes and the genera *Candidatus Solibacter*, *Fimbrimonas*, *Gemmatimonas*, *Sphingobium*, and *Methanobacterium*, were necessary to obtain good fit indices in the SEM. In addition, the composition of the bacterial groups differed depending upon metabolic targets, carbon and nitrogen, and their stable isotopes, $\delta^{13}C$ and $\delta^{15}N$, respectively. Thus, these observations suggested that beetle larvae should incubate partly environmentally beneficial bacteria candidates. Those statistical derived causal structural models highlighted that the larvae fecal enriched bacteria and common symbiotic bacteria might selectively play a role in wood biomass's carbon and nitrogen metabolism. It could confer a new perspective that helps environmental industries in the future.

*Abbreviations*: anammox, anaerobic ammonium oxidization; SDGs, sustainable development goals

## Introduction

Although insects are the most abundant organisms worldwide and account for more than 50% of all known animal species, their biodiversity is threatened, and dramatic rates of decline are predicted [1-5]. Since insects play various roles in ecosystems [6-14] and their importance is evident, many studies on planetary boundaries [15] and sustainable development goals (SDGs) are required [16]. With the increasing need to meet SDGs, the protection of woody biomass is also urgent. Insects coexist by utilizing wood biomass as food and habitat. Since thinning of wood is essential for protecting forests [17], efficient wood recycling by using insects is expected to protect insects and forests by enhancing their ecologically relevant capabilities.

Possible methods for efficient recycling of wood include using wood components themselves as a raw material for biomass energy and the use of insects, which can be cultivated by using wood as food. Insects have various applications [18-23], *e.g.,* for decomposition of wood-derived cellulose to sugar for the production of bioethanol [24], as protein sources in the feed of industrial animals [25], and as feed additives [26] to improve the feed conversion ratio of livestock animals to utilize cellulose within the feed better. To realize using insects in such applications, it is necessary to understand the symbiotic microorganisms associated with digestion in insects. Research on the significance of symbioses between insects, more phylogenetically distant arthropod taxa and their symbiotic microorganisms for the utilization of wood biomass residues has been performed under various viewpoints [27-37]. As the most abundant organisms worldwide, insects may play multiple essential roles in ecosystems. Therefore, studies that aim to understand the potential application of gut microbes in the medical, engineering, and industrial fields and the protection of ecosystems [38,39] are becoming increasingly necessary.

As a first step toward meeting these goals, the bacterial community in larvae of the Japanese rhinoceros beetle (*Trypoxylus dichotomus*), an insect model that utilizes wood, is analyzed by a non-invasive method here. The symbiotic relationships between environmentally beneficial bacterial groups candidates in the habitat of beetle larvae and their excreted feces were investigated by correlation and association analysis with data of total carbon, total nitrogen, and their stable isotopes (Fig. 1). Structural equation models were constructed based on the hypotheses sets using these data. As the results of these calculations, bacterial groups that could play a part in the utilization of wood biomass have been statistically inferred from a bacterial perspective. Furthermore, causal mediation analysis, LiNGAM, and BayesLiNGAM were applied to estimate the importance of microbial groups for the metabolism of carbon and nitrogen and their stable isotopes.

Thus, these observations will infer the role of beetle larvae as a potential incubator for environmentally beneficial bacterial candidates utilizing wood biomass and will allow us to speculate on the systemic relationships of microorganisms that utilize wood biomass from a bacterial point of view. The findings are expected to provide an essential viewpoint that will lead to the conservation of natural components strongly related to insects and innovation in environmental industries in the future.

## Materials and Methods
### Sample preparation
Japanese rhinoceros beetles (*T. dichotomus*) living in a forest in Japan (E140°10′N36°75′) were collected. Thereafter, they were subcultured and naturally mated within the same box (box size: length 33.5 cm x width 49.0 cm x height 31.0 cm) with wood chips (Kunugi mat, Daiso Co., Ltd.) as the habitat. The larvae were subcultured in the same box, covered to prevent adults from flying out, and wood chips were randomly added. After that, mating was repeated from 2015 until 2018. Then, in May 2018, male and female larvae were transferred to the same small box (box size: vertical 13.0 cm x horizontal 21.0 cm x height 13.5 cm), and fresh wood chips were placed in the box (n=2, with n=1 male and n=1 female per box). Since an appropriate moisture content of wood chips is necessary to provide a suitable habitat for beetle larvae, water was moderately applied via spraying. To evaluate the effects of the environmental bacteria, the six boxes were prepared and divided into two groups as follows (Table S1): 1) usually managed group I (three boxes) with a spray application of tap water and 2) group II (three boxes) with spray application of tap water containing an environmental bacterial solution, i.e., an extract from compost with thermophilic Bacillaceae [40-45]. The extract was adjusted to 20% of the total volume of tap water. Water was sprayed *ad libitum* according to the degree of moisture of the wood chips. However, the timing and number of sprays for groups I and II were the same. Fresh wood chips were added once every two weeks, and the decayed chips (similar to soil) and feces were collected after four weeks. Wood chips, decayed chips (M-chips, as indicated in the figures and tables), and feces were used for analyses in this study.

### Bacterial community analysis
DNA from the decayed wood chips and feces was isolated using a QIAGEN QIAamp PowerFecal DNA Kit according to the manufacturer's protocol (QIAGEN, USA). The DNA concentrations were evaluated using the Quant-iT™ PicoGreen dsDNA Assay Kit (Thermo Fisher Scientific, USA). The V4 region (515F-806R) of the bacterial 16S rRNA gene was sequenced on an Illumina MiSeq instrument according to a previous study [46,47]. The obtained sequences were filtered by Trimmomatic (http://www.usadellab.org/cms/?page=trimmomatic).

The 10,000 trimmed reads per sample were analyzed with QIIME 1.9.1 as previously described [48]. The filtered sequences were clustered into operational taxonomic units (OTUs), which were defined at 97% similarity. The α-diversity, β-diversity, bacterial community, and correlations were visualized by using the packages "genefilter", "gplots", "ggplot2", "RColorBrewer", "pheatmap", "ape", "base", "dplyr", "easyGgplot2", "knitr", "ggthemes", "phyloseq", and "vegan" in R software (versions 3.6.2 and 4.0.5) [49,50], Microsoft 365, and Prism software (version 9.1.0). The number of observed OTUs and Chao1, Shannon, and Simpson index values, as measures of α-diversity, were assessed. The β-diversities were estimated by principal coordinate analysis (PCoA) using weighted or unweighted UniFrac distances based on the OTU distribution across samples. The relative abundances of bacterial phyla and genera were selected from the majority (> 1% of the detected population) and represented. The UniFrac distances were analyzed by Adonis in the packages "vegan" and "MASS" of R software, and graphs of α-diversity and UniFrac distances were also prepared by using R software [49,51]. The relative abundances of individual phyla and genera categorized without discrimination of OTUs were compared within the major community (comparison in >1% of the total bacterial community). The bacterial community was analyzed by a paired t test and Mann–Whitney U test/Wilcoxon sign-rank test, as appropriate. The estimation plots were prepared using Prism software. The relative values of dominant and/or characteristic bacteria were visualized by constructing a correlation heatmap after the Pearson correlation coefficient was calculated for the selected bacteria (> 1% of the detected community) using R software. Bacteria showing marked differences ($P$ < 0.2) between targeted groups were selected from each cluster of phyla, genera, and OTUs (> 1% of the detected community). Phylogenetic trees were constructed with MEGA7, MUSCLE, and iTOL (https://itol.embl.de/).

### *Stable isotope analysis of carbon and nitrogen*

The carbon and nitrogen isotopic compositions were determined by DELTA V Advantage (Thermo Fisher Scientific, USA) and Flash EA1112 (Thermo Fisher Scientific, USA), owned by Shoko Science Co., Ltd., Japan, according to the conventional protocol [52-55] with some modifications. In brief, the samples' total carbon and total nitrogen were measured by Flash EA1112 as follows: The oxidation and reduction reactors were heated to 1000 and 680°C, respectively. The carrier gas (He) flow was approximately 100 mL/min. The length of the separation column was 3 m. The oven temperature of the column was set at 35°C. Acetanilide (Kishida Chemical, Japan) was used as a standard. Stable carbon and nitrogen were measured using elemental analyzer/isotope ratio mass spectrometry (EA/IRMS), a unit of Delta V Advantage interfaced with FlashEA 1112, as follows: The oxidation and reduction reactors were heated to 1000 and 680°C, respectively. The carrier gas (He) flow was approximately 100 mL/min. The length of the separation column was 3 m. The oven temperature of the column was set at 35°C. $CO_2$ gas and $N_2$ gas were used as the reference gases to detect stable carbon and stable nitrogen, respectively. The working standards, Vienna Pee Dee Belemnite (VPDB) for carbon ($\delta^{13}C$), included alanine(19.6 ‰), histidine (10.7 ‰), and glycine (33.8 ‰). The working standards, air for nitrogen ($\delta^{15}N$), included alanine (1.58 ‰), alanine (9.97 ‰), and glycine (20.6 ‰). These standards were also provided by Shoko Science Co., Ltd., Japan. The values of chemical indices were visualized by using Prism software.

### *Association analyses*

Association analysis [56-58] is a technically established method in predictive science and is generally applied to understand the relations between components using relative values. It is convenient to apply when there are missing values. Association rules were determined using criterion values for support, confidence, and lift as previously reported [56-58]. In brief, "support" is the probability of X and Y co-occurring in the transaction dataset:

support (X ⇒ Y) = P (X ∩ Y)

The "confidence" of the rule X ⇒ Y is the conditional probability of observing Y given that X is present in a transaction:

confidence (X ⇒ Y) = P (X ∩ Y)/P (X)

The "lift" of the rule X ⇒ Y is the ratio of the support if X and Y are independent:

lift (X ⇒ Y) = P (X ∩ Y)/P (X) P (Y)

Therefore, higher lift values indicate a higher probability of event Y in the case of condition X. The lift values < 1 do not correlate (independent relationship) between X and Y. Therefore, this study adopted a cutoff value of 1.3 as a lift value threshold for association rules. Here, X and Y are represented as the "source" and "target", respectively. The analysis, also called a market basket analysis, was performed with the packages "arules" and "aruleViz" in R software (version 4.0.5) (https://cran.r-project.org). The association analysis parameters were set as support = 0.063, confidence = 0.25, maxlen = 2 and lift > 1.3. The systemic network was rendered by Force Atlas, Fruchterman Reingold, and Noverlap in Gephi 0.9.2 (http://gephi.org). Sankey diagrams [59] for carbon and nitrogen flow were visualized by the packages "networkD3", "tibble", and "tidygraph" in R software. With the bacteria as a source, those directly involved in targeting $\delta^{13}C$, $\delta^{15}N$, carbon, and nitrogen were selected from Figs. 5 and 6. The values in the diagrams were calculated based on the difference in levels of $\delta^{13}C$, $\delta^{15}N$ carbon, and nitrogen between the habitat (M-chips) and the feces (Feces) (Fig. 3) in each box (Table S1). The classification of an increase (_H) and/or a decrease (_L) in the diagrams was also performed in the same way.

### *Structural equation modeling*

Structural equation modeling (SEM) for confirmatory factor analysis (CFA) was conducted using the package "lavaan" [60,61] of R software. The analysis codes were developed with reference to the website (https://lavaan.ugent.be). Since CFA requires a hypothesis, the bacterial groups selected by association analysis were utilized as factors for a latent construct of metabolism of carbon, nitrogen, and their isotopes. The models, serving as hypotheses, were statistically estimated using maximum likelihood (ML) parameter estimation with bootstrapping (n = 1000) by the functions' lavaan' and 'sem'. Model fit was assessed by the chi-squared p-value (p >0.05, nonsignificant), comparative fit index (CFI) (>0.9), root mean square error of approximation (RSMEA) (<0.05), and standardized root mean residuals (SRMRs) (< 0.08) [62]. The path diagrams of the good model were visualized using the package "semPlot" of R software [63].

### *Other statistical analyses for verification of the structural equation model*

Individual causal mediation analyses were performed using the package "mediation" of R software [64]. The analysis codes were developed with reference to the tutorial website (https://rpubs.com/Momen/485122). First, each regression relationship with '~' in the selected model was assessed by using the function 'lm'. Thereafter, the values of the causal relationships between bacterial candidates as mediators and outcomes were evaluated using the function 'mediation'. Finally, the estimated average causal mediation effect (ACME), average direct effect (ADE), and proportion of total effect via mediation were calculated by quasi-Bayesian approximation ('boot=FALSE' as a command) and nonparametric bootstrapping ('boot=TRUE') with

'sims=1000'.

Furthermore, to estimate a structural model beyond the distribution of limited experimental data, the linear non-Gaussian acyclic model (LiNGAM) approach [65], which involves independent component analysis and a non-Gaussian method for estimating causal structures. The LiNGAM was established with Python code on the website (https://github.com/cdt15/lingam) (Python version 3.6). The data calculated by the LiNGAM were visualized as networks in Gephi 0.9.2. The BayesLiNGAM [66], which is a Bayesian score-based approach, were applied for the selected bacterial groups related to the metabolism of carbon and nitrogen. The BayesLiNGAM was established by the package "fastICA" of R software based on the specialized prepared website information (https://www.cs.helsinki.fi/group/neuroinf/lingam/bayeslingam/). According to other website information, the analysis procedure was performed by the package "fastICA" of R software according to other website information (https://cran.r-project.org/web/packages/fastICA). The results of the BayesLiNGAM were visualized as networks by the R package "igraph".

### Statistical analysis

The procedure for statistical analysis was described above for each method. Significance was declared at $P < 0.05$, and a tendency was assumed at $0.05 \leq P < 0.20$. The data are presented as the means ± SEs.

## Results
### Microbial analysis

The fecal bacterial communities of Japanese rhinoceros beetle (*T. dichotomus*) larvae and the wood chips used as their habitat beds were examined. After subculturing the beetles alone and mating them within the same box for two years, the relationship between the bacterial community of the larvae and their habitat was analyzed after the replacement of wood chips and maintenance under stable conditions for one month before the larvae became pupae (Table S1). Despite the presence of different environmental microorganisms, the predominant bacterial community members and bacterial diversities did not appear to change significantly, even if compost was added externally (Experimental conditions I and II in Figs. S1 and S2). Still, the phyla Chloroflexi [67,68], Gemmatimonadetes [69], and Planctomycetes [70] had increased tendency (Figs. 2, S2a, and S3a). When the relative abundances of the genera bacteria were examined in the major community (comparison of bacteria with an abundance >1% of the total bacterial abundance) (Fig. S2b), the markedly enriched bacteria in the feces were as follows (enriched fecal bacteria) (Figs. 2bc, S2b, and S3b): the genus *Bradyrhizobium*, which is a genus of nitrogen-fixing bacteria that affects plant growth and belongs to the phylum Proteobacteria; the genera *Chonella*, *Hyphomicrobium*, and *Paenibacillus*, which are nitrogen-fixing bacteria (rhizobia) [71-73]; the genus *Corallococcus*, members of which have anti-gram-negative activity and act as natural antibiotics [74]; the genus *Gemmata*, which is involved in anaerobic ammonium oxidization (anammox) [69] and belongs to Planctomycetes; the genus *Lutibacterium*, which includes a bacterium that can degrade hydrocarbons and is a candidate for use in bioremediation [75]; *Rhodoplanes*, a genus of bacteria that produce hopanoids [72]. The phylogenetic tree with relative abundances of 15 OTU-assigned bacteria with marked differences (p<0.2) is shown (Fig. S4a). These bacteria were not always phylogenetically closely related to each other. OTU-4121, OTU-3393, and OTU-6, closely related to the genera *Corallococcus*, *Lutibacterium*, and *Rhodoplanes*, were markedly enriched in the feces (Fig. S4b). OTU-4009, closely related to the phylum Verrucomicrobia, was also enriched.

### Correlation analysis of bacterial communities

The phyla Chloroflexi [67,68], Gemmatimonadetes [69], and Planctomycetes [70] were enriched in the feces (Figs. 2 and S3a). However, when the correlations between these bacterial groups in wood chips and feces were investigated, they showed slight correlations in the habitat (Fig. 3a). The genera *Corallococcus* [74], *Gemmata* [69], *Chonella* [71], *Hyphomicrobium* [76,77], *Paenibacillus* [73], *Lutibacterium* [75], and *Rhodoplanes* [72,78], which were enriched in the feces (Fig. 2 and S3b), were not always correlated in the habitat (Fig. 3b). These relationships suggest that increases in bacterial abundance in the feces are independent of the increases in bacterial abundance in the habitat. A positive correlation in the fecal community was shown between the phyla Planctomycetes [70] and Verrucomicrobia, and a negative correlation was shown between the phyla Gemmatimonadetes and Firmicutes (Fig. S5a). The genus *Bradyrhizobium* showed a positive correlation with the genera *Fimbrimonas*, *Gemmatimonas*, *Candidatus Solibacter*, *Pilimelia*, *Hyphomicrobium*, *Pedomicrobium*, *Planctomyces*, *Devosia*, and *Rhodoplanes*. The genus *Chonella* showed a positive correlation with the genera *Paenibacillus*, *Clostridium*, *Cellulomonas*, *Methanobacterium*, and *Pseudonocardia* within the feces (Fig. S5b). The genus *Corallococcus* showed a positive correlation with the genera *Burkholderia*, *Sphingomonas*, *Hyphomicrobium*, *Pedomicrobium*, and *Planctomyces*. The two genera *Corallococcus* and *Lutibacterium* showed a negative correlation with the genus *Optitutus*. The genus *Gemmata* showed a negative correlation with the genus *Candidatus Xiphinematobacter*. The genus *Hyphomicrobium* showed a positive correlation with the genera *Burkholderia*, *Sphingomonas*, *Pilimelia*, *Pedomicrobium*. The genus *Paenibacillus* showed a negative correlation with the genera *Devosia* and *Rhodoplanes*. The genus *Rhodoplanes* showed a positive correlation with the genera *Hyphomicrobium*, *Pedomicrobium*, *Planctomyces*, and *Devosia*. In addition, the patterns of these bacterial relationships in the habitat were different from those of the fecal bacterial relationships (Fig. S6). The abundances of these bacteria in the feces were not necessarily positively correlated with the abundances of bacteria in the habitat.

### Carbon and nitrogen levels and their correlations with the bacterial community

The functional roles of the bacterial community were validated by evaluating the levels of chemical indices, i.e., stable isotopes ($\delta^{13}C$ and $\delta^{15}N$), total carbon, total nitrogen, and the carbon/nitrogen (CN) ratio, in the wood chips and feces. Measurement of the stable isotope contents is known to be useful for inferring metabolic trends in the dietary carbon and nitrogen of insects [23]. The raw chips used as fresh chips for the habitat, decayed chips (M_chips), and feces were examined (Fig. S7a). The levels in both decayed chips and feces were confirmed to be significantly different from those in fresh raw chips (Fig. S7b). In particular, the nitrogen level, $\delta^{15}N$ content, and total N content were clearly higher in the excreted feces of the beetle larvae (Figs. 4a and S7b). As a result, the CN ratio also appeared to decrease in the feces. This tendency changed slightly in M-chips when compost was added externally, but no significant difference was confirmed in excreted feces (Conditions I and II in Fig.S7b). As seen in Fig. 4b, a correlation analysis including the chemical indices revealed positive correlations of total nitrogen with *Lutibacterium*, *Paenibacillus*, and *Chonella*. In addition, the genera *Candidatus* Solibacter and *Sphingomonas* showed weak negative correlations with total nitrogen. The $\delta^{15}N$ content showed a weak negative correlation with the genera *Sphingomonas* and *Coprococcus*. The total carbon content showed weak negative correlations with the genera *Corallococcus*, *Lutibacterium*, *Paenibacillus*, and *Chonella* and a positive correlation with the genus *Candidatus Solibacter*. The $\delta^{13}C$ content showed a positive correlation with the genera *Lutibacterium* and *Elin506* and a negative correlation with the genera *Burkholderia*, *Devosia*, *Sphingomonas*, *Pilimelia*, and *Fimbrimonas*. The CN ratio showed negative correlations with the genera *Cellulomonas*, *Pseudonocardia*, and *Methanobacterium*.

### Predictive selection of the bacterial community for carbon-nitrogen metabolism

Association analysis [56-58], a technically established method in predictive science, was applied to evaluate the correlations of chemical indices and microbial communities in excreted feces beyond the bacterial taxonomic boundaries of the genus and phylum levels. The results showed that the systemic network of the indices and the bacteria in decayed wood chips and feces could be classified into four categories (Fig. S8). Based on the calculated lift values, the networks among the indices and the bacteria enriched in the feces are represented in Fig. S9. The associations of the bacterial genera *Corallococcus*, *Gemmata*, *Hyphomicrobium*, and *Lutibacterium* increased in the feces. The other enriched and unenriched fecal bacteria were linked via feces-associated bacteria and/or other bacteria (Figs. S9 and S10). The chemical indices were strongly linked with the bacteria enriched in the feces and bacteria that showed an inverse correlation with the indices. $\delta^{13}C$ was connected with the phylum Acidobacteria, which was mediated with the genus *Elin506*, which in turn was linked with enriched fecal bacteria. The analysis that identifies high and low as levels estimated that a high level of $\delta^{13}C$ was modulated by a group related to the increased phylum Acidobacteria and decreased phylum Armatimonadetes and genus *Sphingobium* (Fig. 5a). A low level of $\delta^{13}C$ was modulated by a group related to the reduced phylum Acidobacteria and increased phylum Armatimonadetes and genera *Bradyrhizobium*, *Fimbriimonas*, *Methanobacterium*, and *Sphingobium* (Fig. 5a). $\delta^{15}N$ was linked with *Gemmata*, which was enriched in the feces, and the genera *Fimbrimonas* and *Methanobacterium*, which were not constantly enriched there (Fig. S9). A high level of $\delta^{15}N$ was modulated by a group related to the increased phylum Armatimonadetes and genera *Gemmata*, *Fimbrimonas*, and *Methanobacterium* (Fig. 5b). A low level of $\delta^{15}N$ was modulated by a group related to the decreased phylum Armatimonadetes and genus *Gemmata* (Fig. 5c). The total carbon content was linked with *Bradyrhizobium*, one of the enriched fecal bacteria, and the phylum Bacteroidetes and genus *Candidatus Solibacter*, which were connected with some of the enriched fecal bacteria (Fig. S9). In particular, high total carbon content was modulated by a group related to the increased genera *Bradyrhizobium* and *Candidatus Solibacter* (Fig. 6a). Low carbon content was modulated by a group related to the decreased genera *Bradyrhizobium* and *Candidatus Solibacter* (Fig. 6b). The total nitrogen content was linked with the genus *Gemmatimonas*, which was connected with the genus *Pilimelia*, which was mediated with enriched fecal bacteria (Fig. S9). High total nitrogen content was modulated by a group related to the increased genus *Gemmatimonas* and decreased genera *Devosia*, *Optitus*, and *Sphingobium* (Fig. 6b). Low total nitrogen content was modulated by a group related to the reduced genus *Gemmatimonas* and increased genera *Devosia*, *Optitus*, *Methanobacterium*, and *Sphingobium* (Fig. 6a). The CN ratio was linked with the enriched fecal bacteria via the genera *Devosia*, *Methanobacterium*, *Optitus*, and *Sphingobium*, minor members of the fecal bacterial community (Fig. S9). A high CN ratio was modulated by a group related to the decreased genus *Gemmatimonas* and increased genera *Devosia*, *Methanobacterium*, *Optitus*, and *Sphingobium* (Fig. 6c). A low CN ratio was modulated by a group concerning the increased genus *Gemmatimonas* and decreased genera *Devosia*, *Optitus*, and *Sphingobium* (Fig. 6d).

These observations indicated the possibility that the metabolism of carbon and nitrogen was associated with the enriched fecal bacteria and other bacteria that were minorities in the fecal bacterial community and were inversely correlated with these indicators.

### Structural equation modeling of the bacterial community for carbon-nitrogen metabolism

Since bacterial groups involved in carbon and nitrogen metabolism could be predicted by association analysis, structural equation models based on these hypotheses were established. As a result, we constructed relatively ideal structural models for the goodness-of-fit indices (Tables S2), which could be analyzed by bootstrapping using the maximum likelihood method even when the number of samples was relatively small. First, a bacterial group associated with stable isotope carbon $\delta^{13}C$ selected by the association analysis (Fig. 5a) was statistically tested as hypothesized factors. A regression group with the phyla Acidobacteria, Armatimonadetes, the genera *Bradyrhizobium,* and *Sphingobium* was a more suitable structural model with the highest goodness of fit for stable isotope carbon metabolism (No.1 of $\delta^{13}C$ in Table S2) (Figs. 7a) than other models (the No.2 in Table S2). An appropriate model for the metabolism of stable isotope nitrogen $\delta^{15}N$ was confirmed within the group shown in Fig. 5b by the similar procedure, and a regression group with *Gemmata*, *Fimbrimonas*, *Methanobacterium*, Armatimonadetes, Planctomycetes, and Gemmatimonadetes was selected (No.1 of $\delta^{15}N$ in Table S2)(Fig. 7b) as preferable to other models (the No.2 in Table S2). Suitable models for total carbon and nitrogen metabolism were selected (No.1 of Total C and Total N in Table S2)(Figs. 7c) over other models (the No.2 in Table S2). The appropriate model of total carbon metabolism did not necessarily match the suitable model of stable isotope carbon metabolism (Figs. 7ac), and Bacteroidetes was present in the structural model without *Bradyrhizobium* and *Candidatus Solibacter*. The appropriate model of total nitrogen metabolism was inconsistent with the suitable model of stable isotope nitrogen metabolism (Figs. 7bd). Among bacteria in the suitable model for the metabolism of total nitrogen, *Bradyrhizobium*, *Sphingobium,* and/or *Candidatus Solibacter* were also involved in models for the metabolism of stable isotope carbon and/or total carbon. *Paenibacillus*, *Optitus*, *Devosia*, and *Gemmatimonas* were characteristic of a total nitrogen metabolism model. The common bacteria that were not significantly different between M-chips and feces were necessary in the suitable models of carbon and nitrogen metabolism. Finally, the causal relationship in the bacterial group that had a regression relationship according to SEM was statistically confirmed by causal mediation analysis (Tables S3-S6). As a result, it became clear that individual bacteria do not necessarily have causal relationships, with some exceptions. In particular, the genus *Sphingobium* and the phylum Armatimonadetes, minor bacteria, appeared to play an essential role in the metabolism of stable isotope carbon $\delta^{13}C$ as shown in Table S3 (p=0.03799 and p=0.0235, respectively). The presence of the genera *Fimbriimonas* and *Metanobacterium* and the phylum Armatimonadetes, minor bacteria, appeared to play a role in the metabolism of stable isotope carbon $\delta^{15}N$, although not significantly (Table S4). The presence of the genus *Bradyrhizobium*, which was significantly increased in the feces, appeared to play an essential role in the metabolism of total carbon (p=0.0437) (Table S5). The presence of the genera *Corallococcus* and *Paenibacillus*, which were significantly increased in the feces, appeared to play an essential role in the metabolism of total nitrogen (p=0.0442 and p=0.0557, respectively)(Table S6). The genera *Gemmatimonas*, *Opitutus*, *Candidus Solibacter,* and *Sphingobium*, which are minor bacteria, appeared to play a role in the metabolism of total nitrogen. However, the effect was not always significant.

Regardless, structural equations with a high goodness of fit could be used to formulate such a group by using association analysis results as a hypothesis. Based on the observations, groups for carbon and nitrogen flows are visualized by Sankey diagrams (Fig.8). Thus, these observations suggest that the bacteria involved in carbon and nitrogen metabolism should be affected not by specific predominant bacteria but by an influential bacterial group forming a particular structure.

## Discussion

The findings of this study suggested that potentially environmentally beneficial bacteria selectively grow in the intestines of beetle larvae. Their bacterial populations of beetle larvae are unlikely to change depending on the difference of environmental conditions in the habitat. The phenomenon in which competition occurs among the intestinal bacteria of insects is interesting, although the relevant study did not consider beetle larvae [79]. A statistical method that can be verified even with a small number of samples is used with 1000 times of bootstrapping. Therefore, this observation may have found one of universality in a restricted environment. As the investigations, the interactions between the predominant bacteria and other minor bacteria may regulate carbon and nitrogen flow from wood chips. Other symbiotic microorganisms and the hosts themselves may have affected the metabolism of carbon and nitrogen[39]. However, focusing on bacteria as the subject of research, the patterns as the bacterial group were statistically evaluated to explore a novel rule for wood biomass processing. Although the pursuit of the function of a single microorganism is prioritized, this study suggested that bacterial groups with a non-dominant bacterium may be essential by the statistical methods. It was possible to infer an influence of the group of symbiotic bacteria, not an effect of only a specific bacterium. The metabolic function as an individual bacterium was grasped as follows. For example, these bacterial candidates, the phylum Chroloflexi [67,68], members of which are known to have a nitrogen-oxidizing function [80], were detected by examining relative abundances. Bacteria belonging to the phyla Planctomycetes [70] and Gemmatimonadetes [69] potentiate the regulation of anaerobic ammonium oxidation (anammox) [69,70]. It should be notable that *Gemmata*, a bacterial genus that belongs to Planctomycetes, also participates in anammox [69], which reduces the production of nitrous oxide, $N_2O$[81], a global warming gas[82]. In recent years, $N_2O$ production from the agricultural field has been regarded as a crucial problem because it is known that it has a global warming coefficient about 300 times that of $CO_2$. Based on these backgrounds, the feces of beetle larvae here were expected to suppress the generation of nitrous oxide potentially. In addition, the genera *Chonella*, *Hyphomicrobium*, and *Paenibacillus* are nitrogen-fixing and/or nitrogen-denitrifying bacteria (rhizobia) [71-73,76,77]. *Rhodoplanes*, a phototrophic genus that produces hopanoids [72,78], which are involved in controlling plant root growth and nitrogen fixation, are recognized as plant growth-promoting rhizobia (PGPRs). Among these findings, the presence of nitrogen-fixing bacteria in beetle larvae is consistent with previous data for other insects, such as termites (*Termitidae*) [20-22] and stag beetles (*Lucanidae*) [34], although the species of nitrogen-fixing bacteria were different from those in previous studies. The presence of bacterial candidates with nitrogen-fixing, anammox, and hopanoid-producing abilities was shown in this insect. Furthermore, the genus *Lutibacterium*, which includes a bacterium that can degrade hydrocarbons, is a candidate for use in bioremediation [75], and the genus *Corallococcus*, which has anti-gram-negative activity act as natural antibiotics [74], were enriched under these conditions. Thus, the experiment conducted in this study showed enrichment of bacterial candidates with potential roles in plant-animal-environment symbiosis. These bacteria from beetle larva feces may be involved in nitrogen fixation as PGPR, reduction of global warming gas by anammox reactions, hopanoid production, anti-gram-negative interactions, and hydrocarbon degradation.

Several evaluation methods inferred the symbiotic relationships of these bacteria and their functional roles. Although correlation analyses showed positive correlations between the total nitrogen content and *Chonella*, *Lutibacterium*, and *Paenibacillus* (Fig. 4b), association analyses predicted the relationships of chemical indices via the common bacteria which were inversely correlated with the enriched fecal bacteria. The total carbon content, CN ratio, $\delta^{15}$N content, and $\delta^{13}$C content appeared to be associated with the other bacteria, which were minor members of the fecal bacterial community detected under the experimental conditions in this study. These results suggest that the balance of the abundance ratio of dominant and inferior strains in the intestinal flora of beetle larvae may affect the metabolism. To the best of our knowledge, the SEM interaction network of candidate bacterial groups involved in the carbon and/or nitrogen cycle in the feces of beetle larvae has been statistically estimated for the first time. SEM suggested that different groups regulated the metabolism of carbon, nitrogen, and their stable isotopes. However, causal mediation analysis revealed that the estimate is not significantly important as a single effect. To speculate the spatial-temporal relationship of the bacterial groups, BayesLiNGAM was performed. The percentages of the top 6 patterns of causal relationships were low, although various patterns of direct causal relationships were calculated (Figs. S11). However, the predominant direct causal relationship was not always observed in bacterial groups for the metabolism of $\delta^{13}$C, $\delta^{15}$N, total carbon, and total nitrogen. The results suggest the spatial-temporal relationship of bacterial groups for the carbon and nitrogen metabolism and their complexity.

Since the experimental data in this study were limited, it was necessary to analyze them using the maximum likelihood method based on a Gaussian distribution. However, whether original data follow a Gaussian distribution is not always clear. Therefore, assuming a non-Gaussian distribution, we analyzed the network with LiNGAM [65] as an independent component analysis. Bacteria for the metabolism of $\delta^{13}$C, $\delta^{15}$N, total carbon, and total nitrogen with good fits in SEM were partly selected as the LiNGAM causal networks (Fig. S12). The set components of the causal networks may strongly show causal relationships. Generally, LiNGAM is suitable for the data may not inherently follow a Gaussian distribution in nature [65]. Such network characteristics can be manifested under conditions with non-Gaussian distributions, depending on the environmental conditions.

This study evaluated the relationship between bacteria at the metabolic level under restricted experimental conditions and beetle larvae, but the modeling procedure in this study may be necessary. The bacterial behavior observed in this study may indirectly represent interactions with unknown factors, including other microorganisms. Future evaluations of the relationship with other symbiotic microorganisms, such as fungi and the host itself, will be made. It is expected that the findings of such research will lead to a comprehensive understanding of the symbiotic system and will help study fermentation conditions with wood in an engineering manner.

Predicting the groups of microbes involved in carbon and nitrogen flow may be necessary for future studies. This study may provide a new perspective on diversity in ecosystems and insight for industrial utilization. In particular, the increasing trend of $\delta^{15}$N content in the feces may indicate that nitrogen from the wood chips was utilized for nutrition, as $\delta^{15}$N is abundant in nitrogen obtained from plants [83]. Since the utilization of aerial nitrogen by stag beetle larvae was suggested [34], the nitrogen cycle of species other than beetles may be regulated through various steps. For example, the nitrogen cycle-regulated bacterial group present in the larvae may be involved in protein synthesis. Concerning other observations, nitrogen cycle-regulated bacteria have been suggested to be abundant in the intestines of indigenous people who eat only plants and are muscular [84]. The relation to this report is interesting, although it must be considered with caution because the animal species are different.

Generally, insects display physiological differences concerning metabolism among growth stages [23]. In addition, the microbial community may change depending on symbiotic microorganisms and environments [11,85]. It has been pointed out that the microbial community of beetles may vary depending on the symbiotic microorganisms and the environment [9,10,12,86]. Therefore, it should be noted that our observations in this study were based on a restricted environment just before the larvae became pupae. An essential point of this study is that it was possible to classify the groups of microorganisms with metabolic functions, including potential association groups controlling carbon and nitrogen flow from a bacterial point of view. Although the mechanisms were not clarified in this study, the certainty of the importance of these groups, which are involved in carbon and nitrogen flow, will be revealed in future studies. This is expected to influence the role of insects in the ecosystem. From the perspective of ecosystem maintenance and industry, these observations remind us of the need for research with the following views: The importance of the potential role of humic soil for beetles in the natural cycle, although rarely discussed to date, and environmental functions in future industrial applications (Fig. 1), *i.e.,* in circulating agriculture [87], the prevention of plant and wood loss [88,89], animal health [25], and the prevention of global warming [90] as the above described. The finding that nitrogen-fixing and hopanoid-producing bacteria, which are helpful for plants, were enriched in the feces of larvae-fed wood is significant. These beneficial bacteria in insect feces may be associated with the importance of humic soil, a compost used since ancient times. The enrichment of specific bacteria with natural antibiotics could be helpful in the development of a novel feed additive to replace artificial antibiotics. In addition, the enrichment of specific bacteria with anammox and hydrocarbon degradation abilities could be helpful for bioremediation and environmental protection.

Thus, symbiotic bacteria from beetle larvae may be essential for agricultural recycling and ecological restoration and conservation, and reassessment of the functional grouping of candidate bacteria involved in carbon and nitrogen metabolism can be helpful as a perspective on intestinal metabolism and environmental control. To build a sustainable society, attempts to understand the mechanisms underlying the interactions and roles of environmentally symbiotic microbes of insects will be required.

### Acknowledgments

We are grateful to Mrs. Miki Asakura, Mrs. Naoko Tachibana, Mrs. Sayo Suzuki, and Mrs. Naoko Atarashi (Riken IMS) for providing technical advice and support. In addition, we would like to express special thanks to Mrs. Izumi Jouzuka for helping with figure design.

### Author contributions

Conceived and designed the experiments: H.M., K.I., and M.H.; performed subculturing and sampling: H.M. and K.I.; performed DNA extraction and the NGS experiment: F.A., W.S., M.M., C.S., and T.K.; analyzed the data: N.T., W.S., T.N., A.T., C.I., T.K., A.K., and H.S.; contributed reagents/materials/analysis tools: H.M., K.I., W.S., A.T., T.K., H.S., M.H., A.K., H.O., J.K., and H.K.; wrote the paper: H.M. and J.K.; revised the manuscript: H.M., H.S., S.M., M.H., M.H., H.O., J.K., and H.K.

### Data availability

All 16S rRNA gene datasets were deposited in the DDBJ Sequence Read Archive (accession number: DRA012158). In addition, the R protocols for association analysis used in this study were deposited on the following websites: http://dmar.riken.jp/Rscripts/ and http://dmar.riken.jp/NMRinformatics/.

### Competing interests

The authors declare no competing interests.

**Figure legends**

Fig. 1
Experimental design in this study. As a non-invasive method, bacterial populations in the habitat (wood chips) and feces of beetle larvae and the concentrations of total carbon, total nitrogen, and their stable isotopes were analyzed. First, correlation analysis and association analysis were carried out with these data. Next, hypotheses were made based on the investigation, and then they were verified by the covariance structure equation for confirmatory factor analysis (CFA).

Fig. 2
(a) Relative abundances of phyla in the microbiota in the habitat chips of beetle larvae (bed for larvae) (M-chips) and their feces (Feces)
(b) Differences in phylum- and genus-level communities between the habitat (M-chips) and feces (Feces) (n = 6; $p<0.2$; >1% as the maximal value of the detected bacterial communities among the whole community in each group). * indicates $p < 0.05$.
(c) Estimation plots of representative phyla and (b) genera in Fig. 2b with their significance values ($p<0.1$; >1% as the maximal value of the detected bacterial community among the whole community in each group).

Fig. 3
Heatmaps of correlations between the habitat and fecal microbiota of beetle larvae: (a) phyla and (b) genera.

Fig. 4
(a) Estimation plots of stable isotope ($\delta^{13}C$ and $\delta^{15}N$) levels, carbon and nitrogen levels, and carbon/nitrogen ratios in the decayed chips (M-chips) and larval feces (Feces) represented in Fig. S7
(b) A heatmap of correlations between the components and the genera in beetle larvae feces is shown in Fig. 2b.

Fig. 5
Interactive systemic networks of factors associated with chemical indices and feces, which are shown with high (_H) or low (_L) levels based on the mediation values of the whole data of targeted components: (a), $\delta^{13}C$; (b) and (c), $\delta^{15}N$. The bacteria and components in Fig. 2b and Fig.4a are shown in bold letters. Bacteria with low abundances (Fig.S10) are shown in violet.

Fig. 6
Interactive systemic networks of factors associated with chemical indices and feces, which are shown with high (_H) or low (_L) levels based on the mediation values of the whole data of targeted components: (a)(b), total carbon and nitrogen; (c)(d), CN ratio. The bacteria and components in Fig. 2b and Fig.4a are shown in bold letters. Bacteria with low abundances (Fig.S10) are shown in violet.

Fig. 7
The relationship of the fecal microbiota associated with (a) $\delta^{13}C$, (b) $\delta^{15}N$, (c) total carbon content, and (d) total nitrogen content is shown by structural equation modeling for selected groups in Fig. 5. In addition, standardized β coefficients are reported. The abbreviations are as follows: Brd, Bradyrhizobium; Sph, Sphingobium; Acd, Acidobacteria; Arm, Armatimonadetes; Fmb, Fimbriimonas; Gemmt, Gemmata; Gmmtm, Gemmatimonadetes; Pln, Planctomycetes; Mth, Methanobacterium; Bct, Bacteroidetes; C.S, *Candidatus Solibacter*; T_C, Total_C; Crl, Corallococcus; Dvs, Devosia; Opt, Opitutus; Pnb, Paenibacillus; T_N, Total_N. Green positive; red, negative. The fit indices are shown in Table S2.

Fig. 8
Carbon and nitrogen flow calculated by structural equation modeling in Fig.7 visualized by Sankey diagrams: (a) $\delta^{13}C$, (b) $\delta^{15}N$, (c) total carbon content, and (d) total nitrogen content.

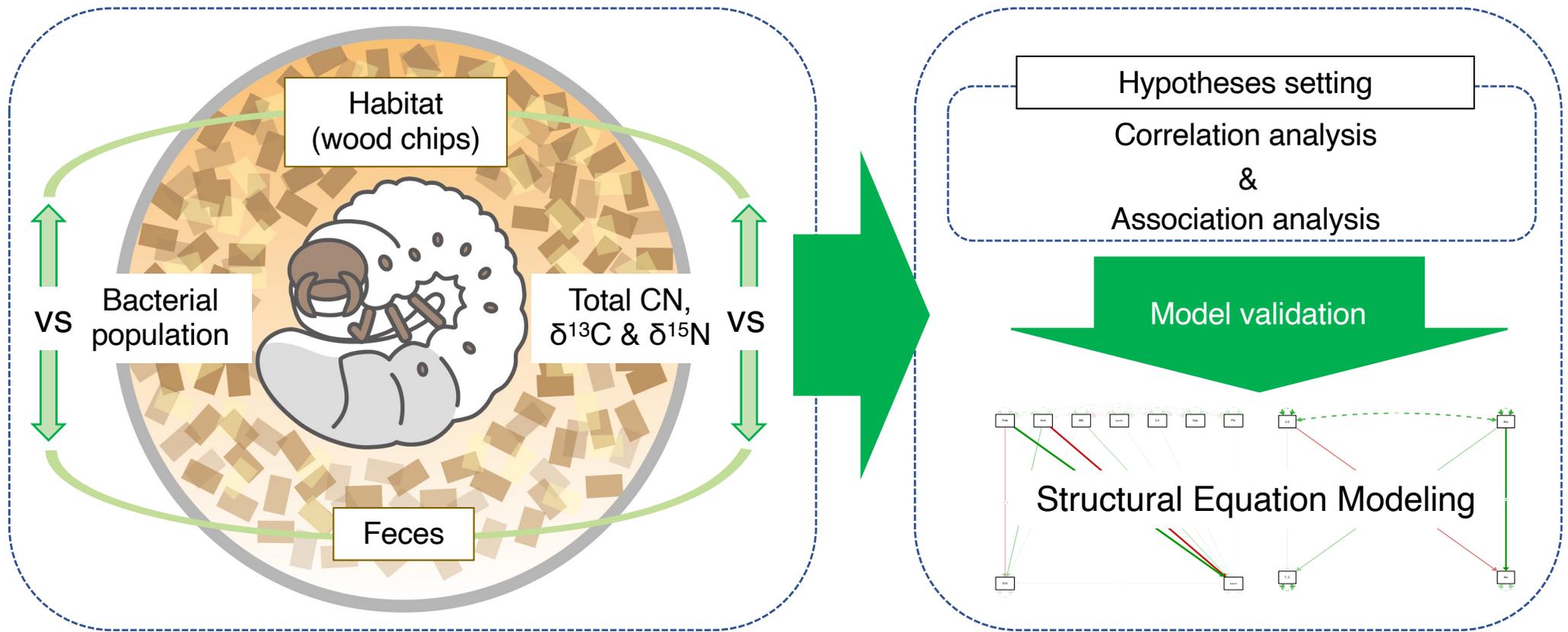

Fig.1

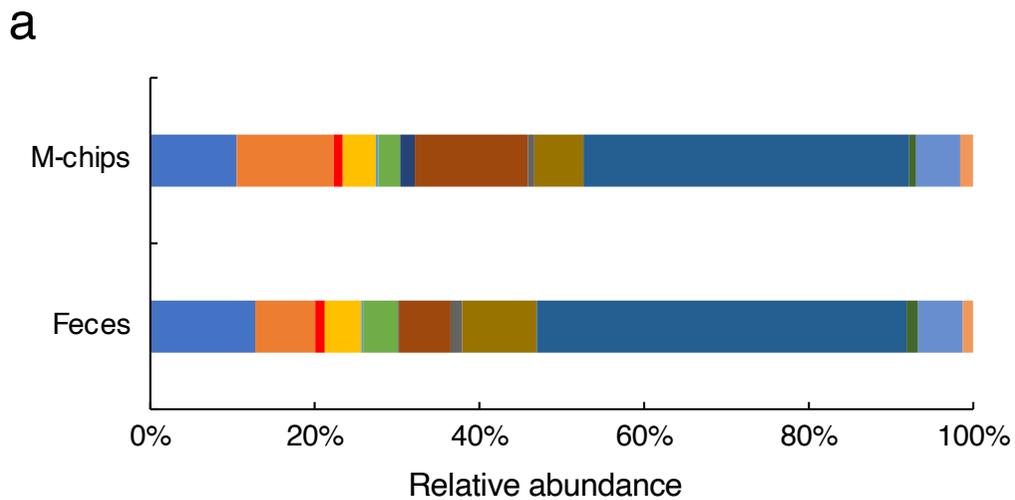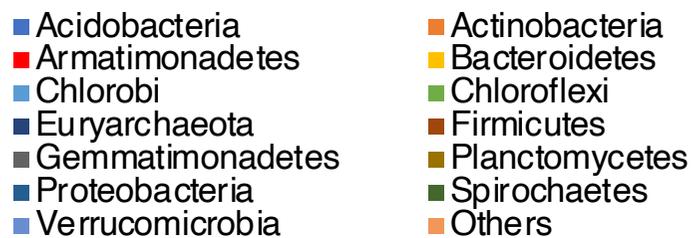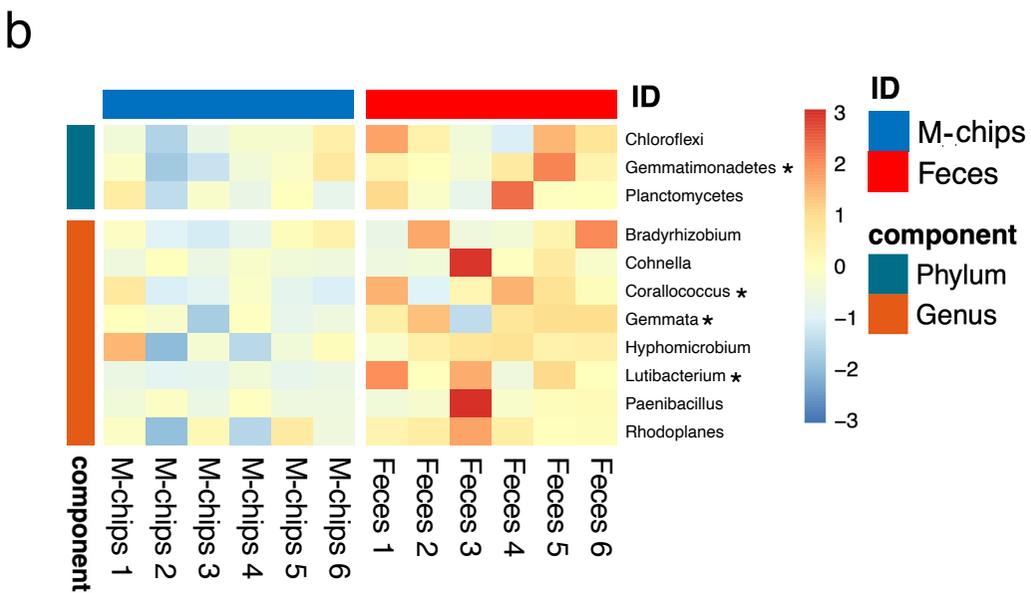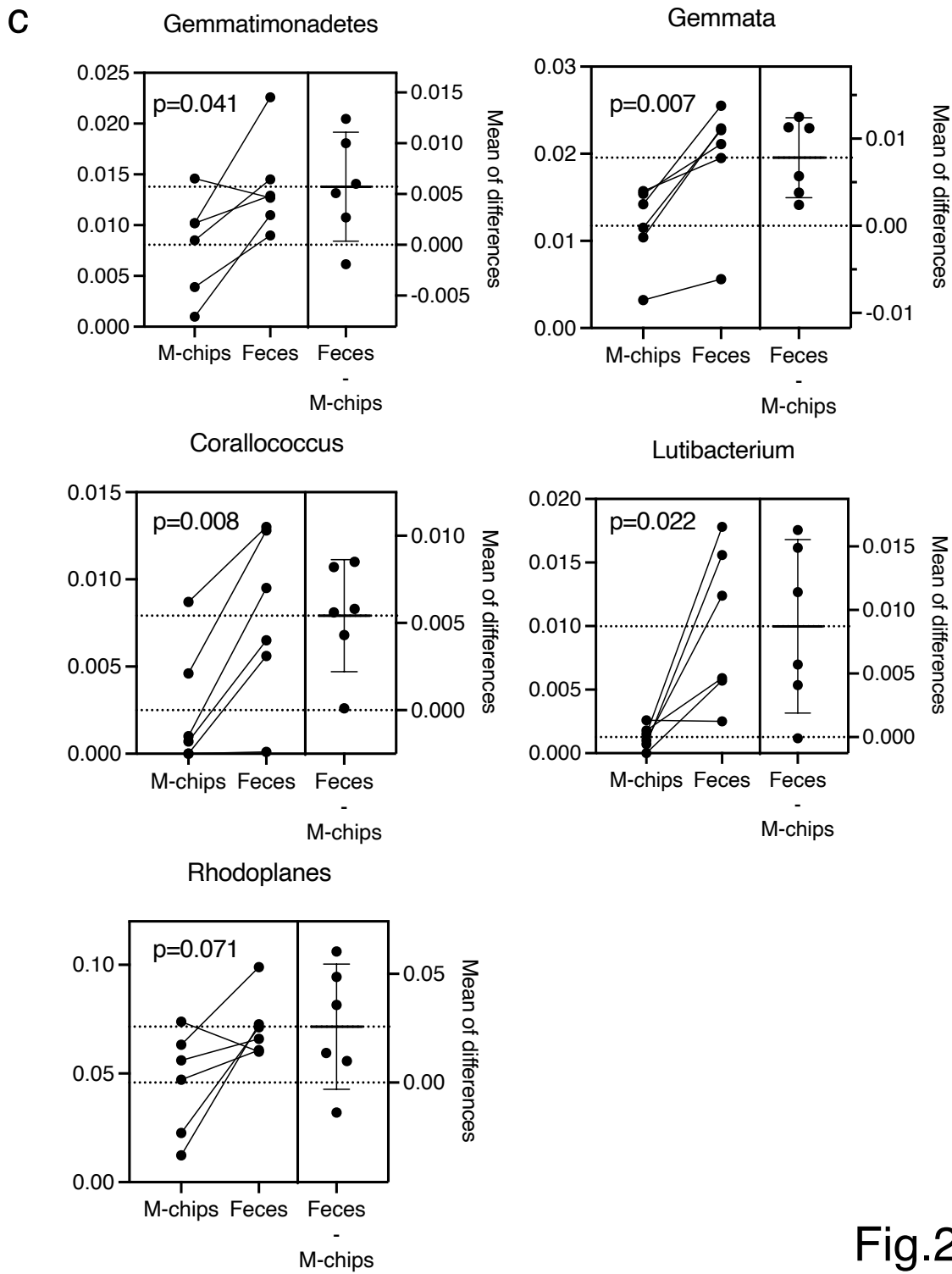

Fig.2

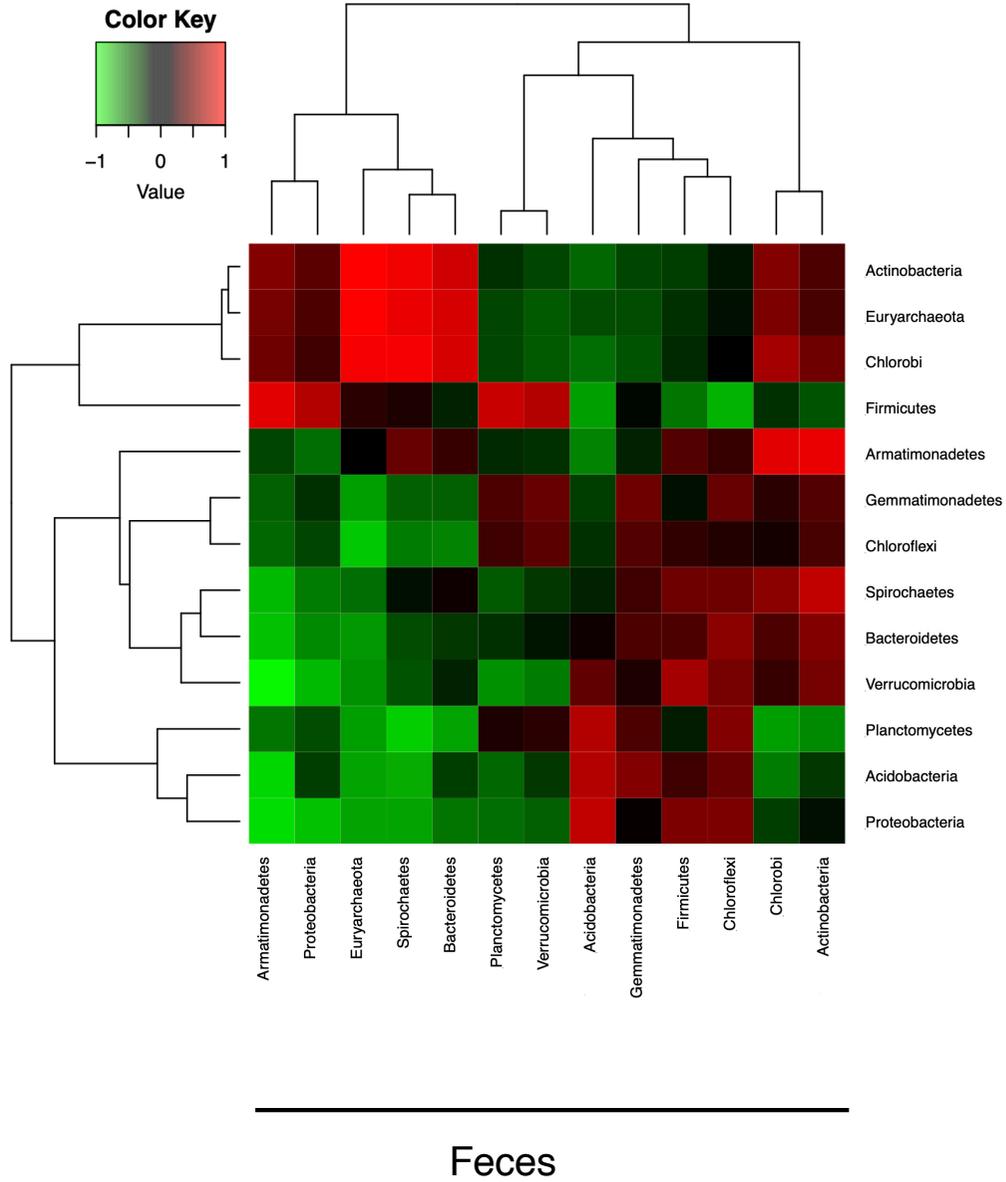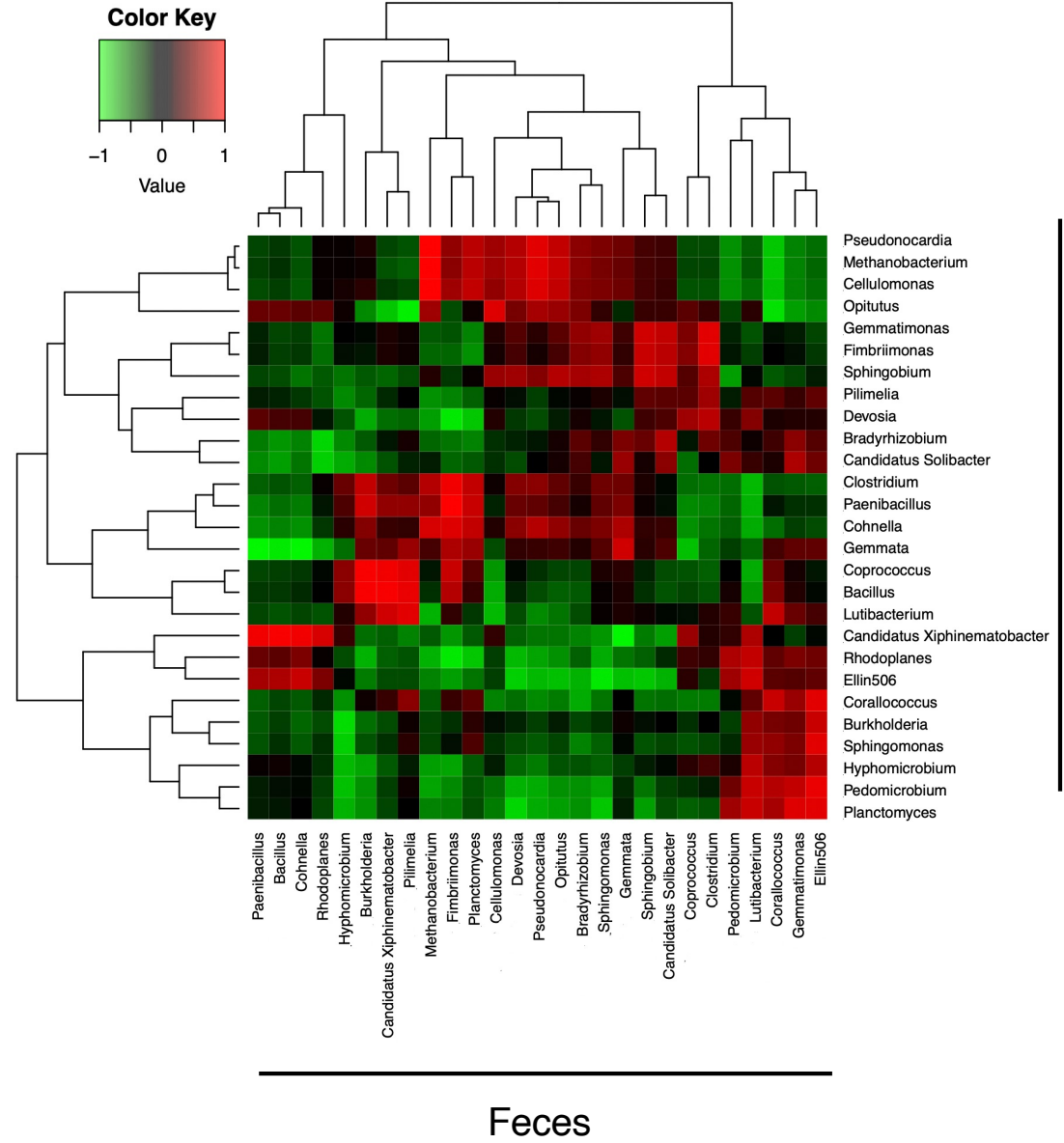

Fig.3

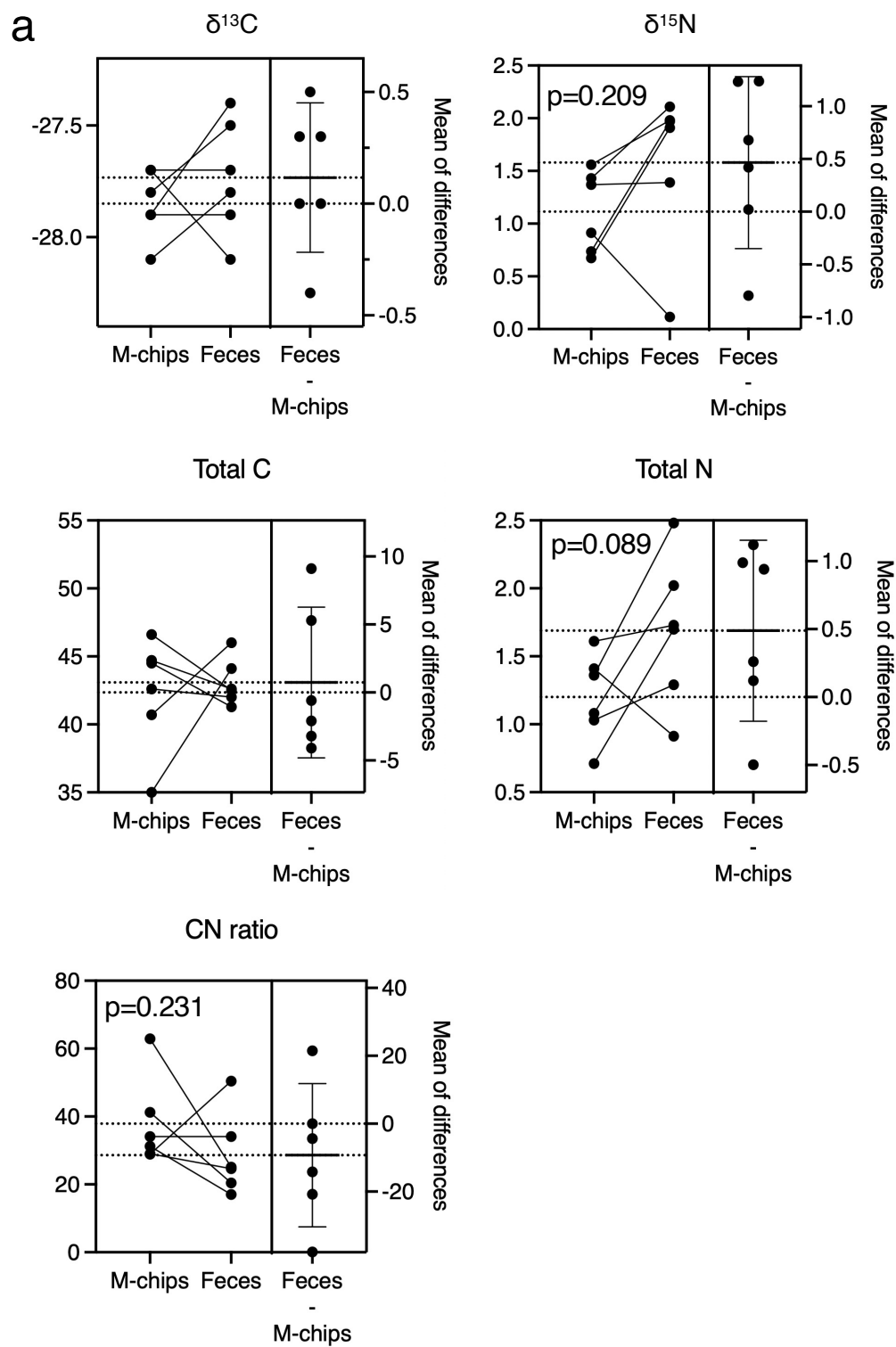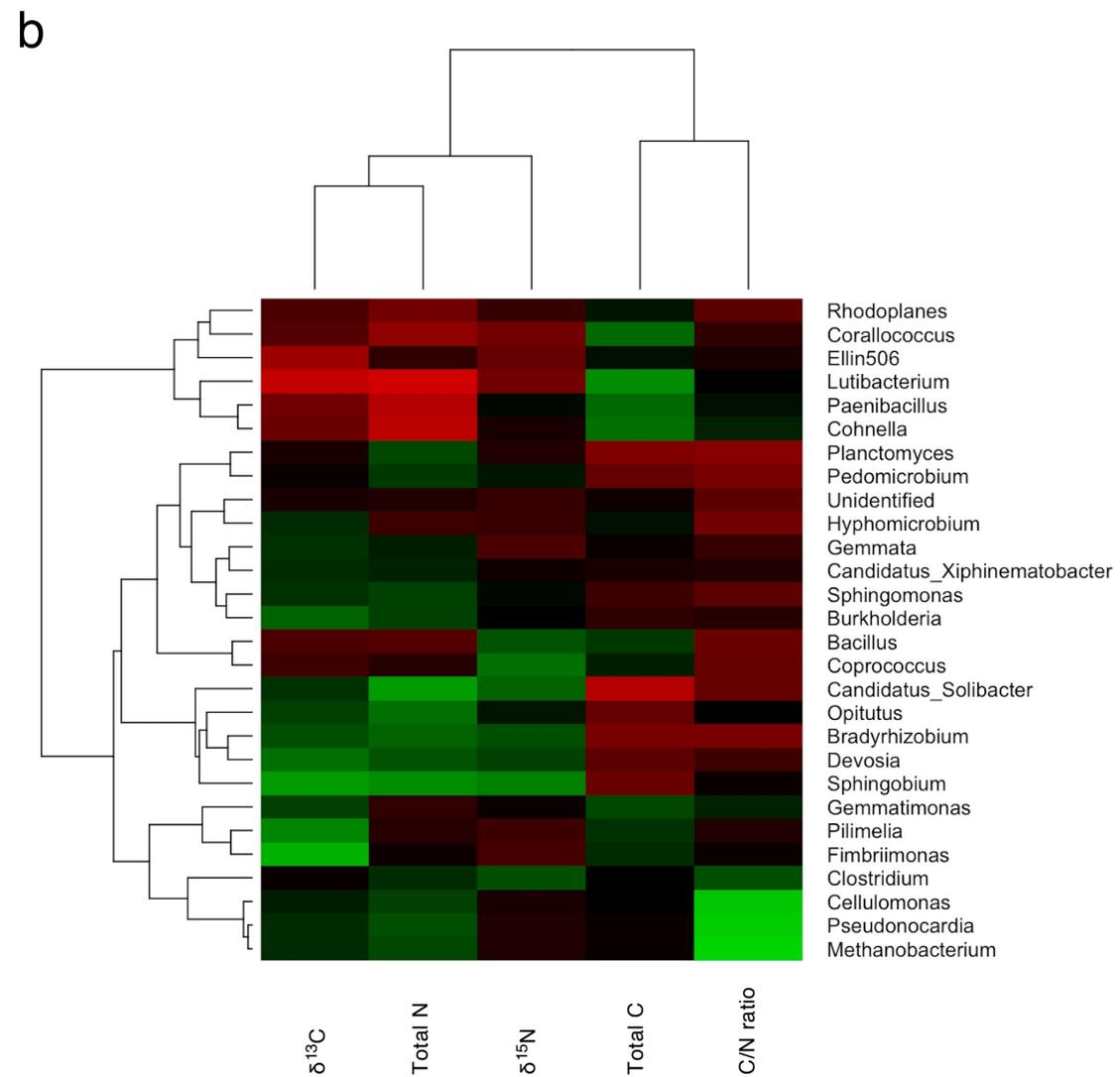

Fig.4

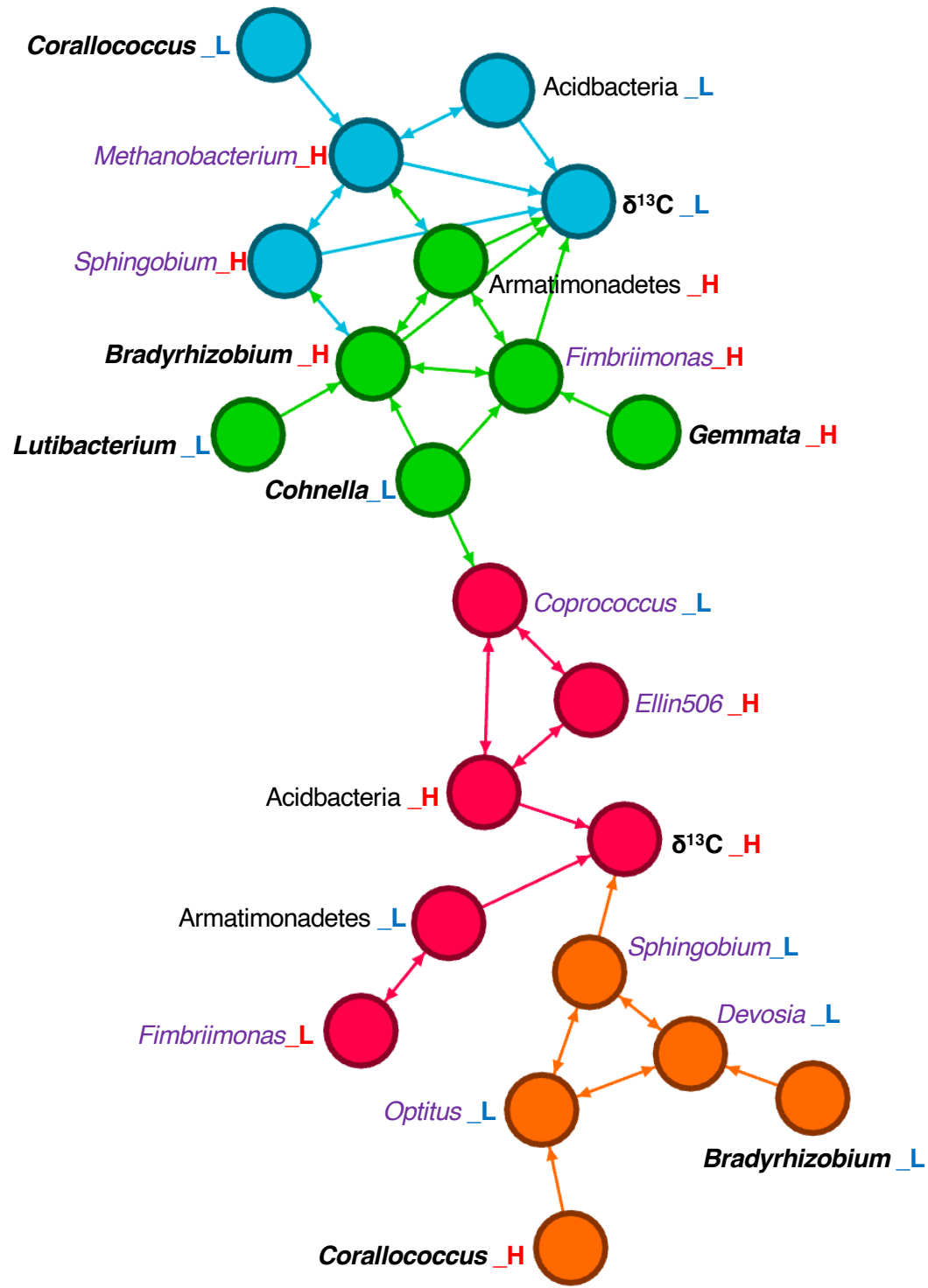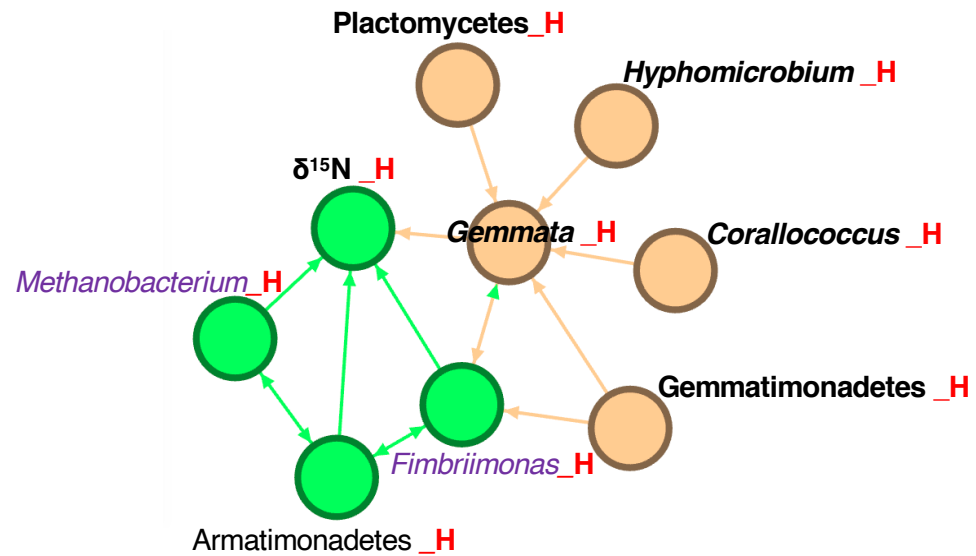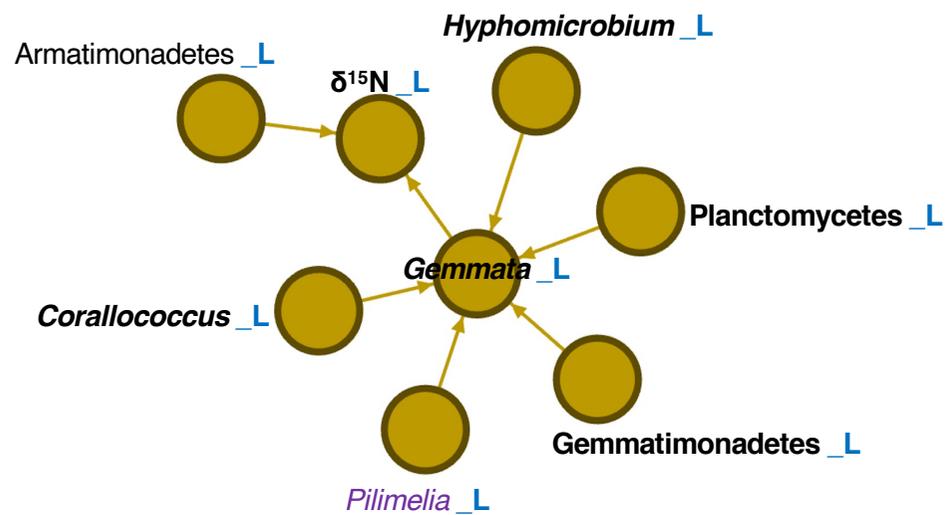

Fig.5

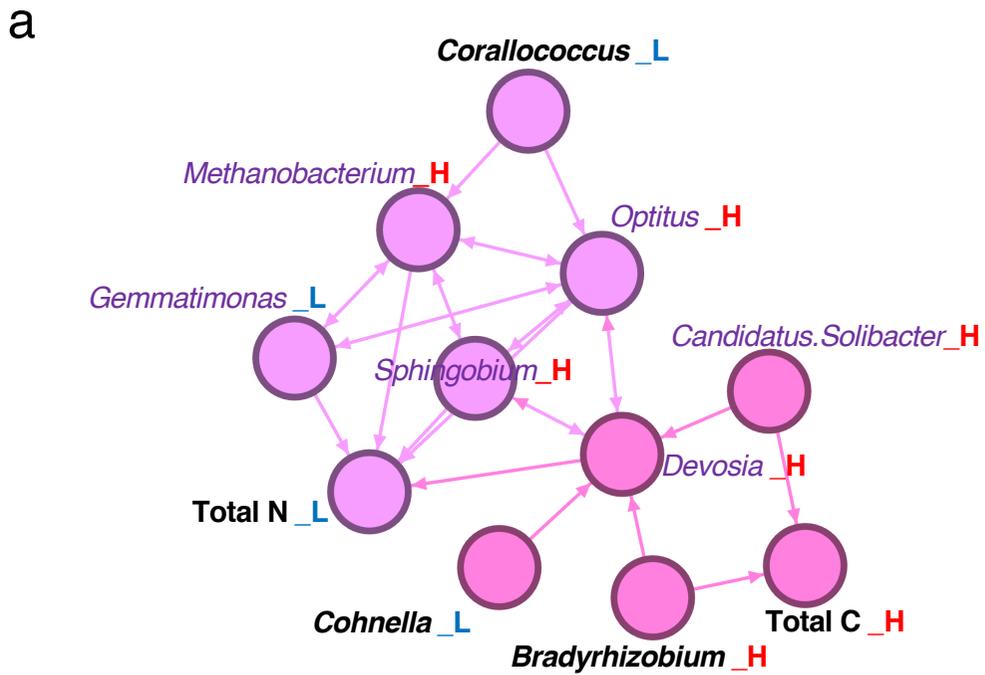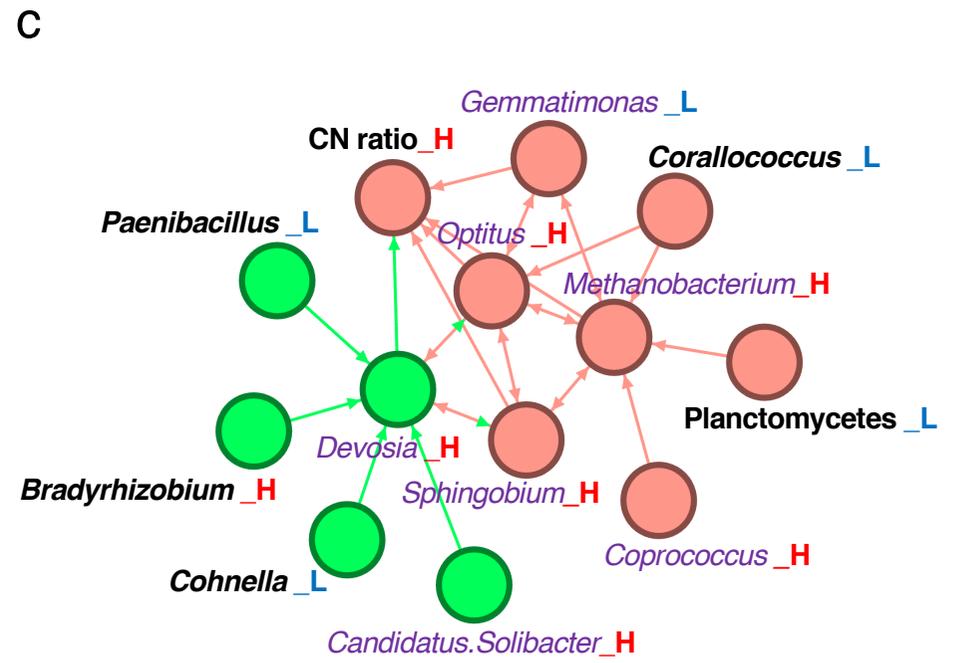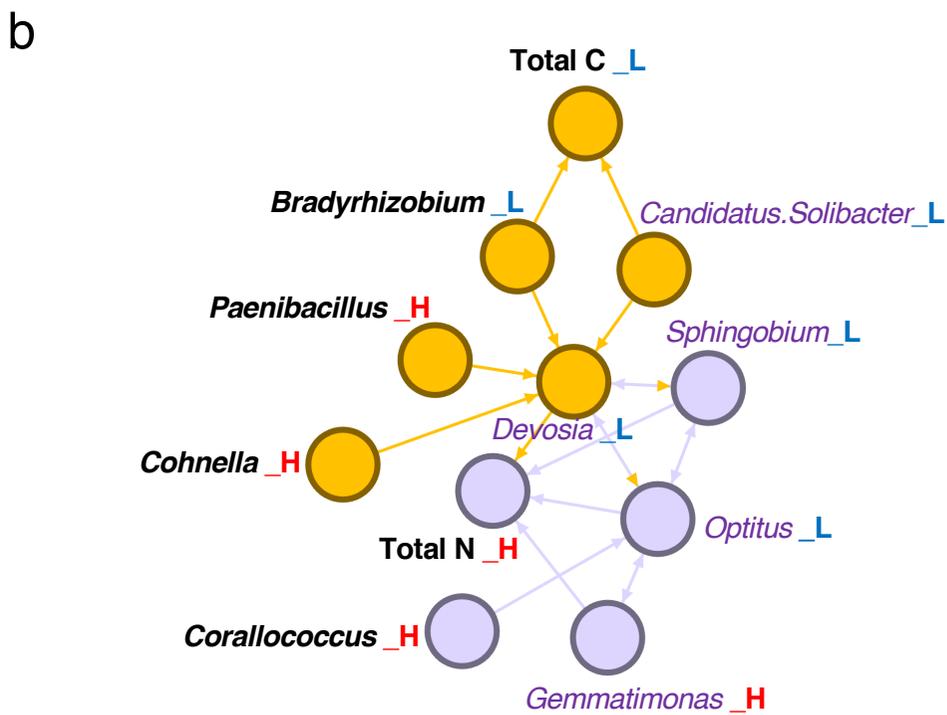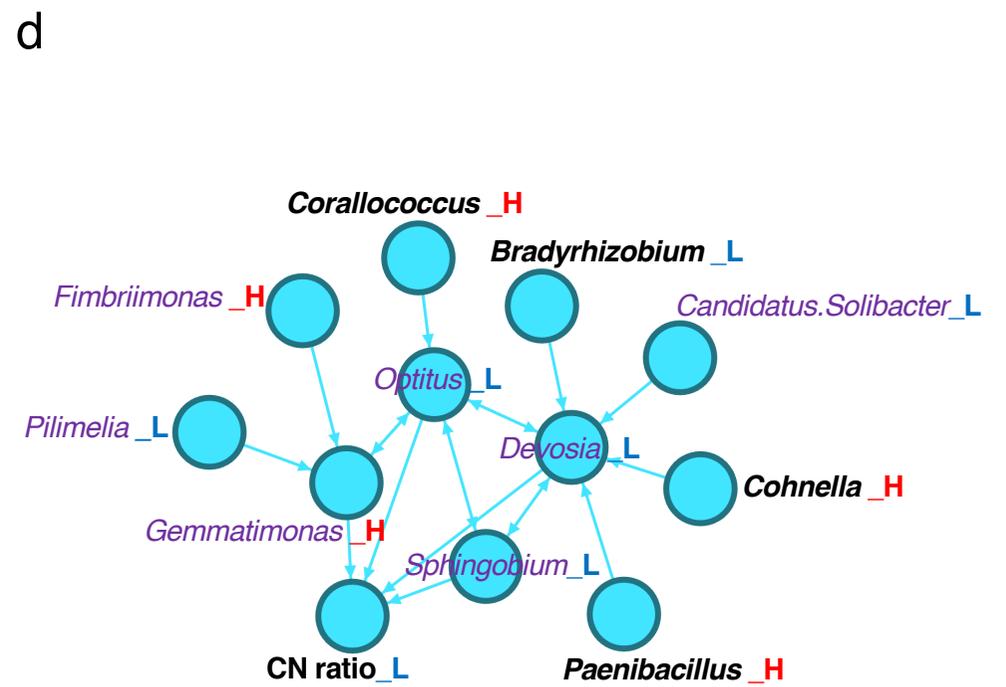

Fig.6

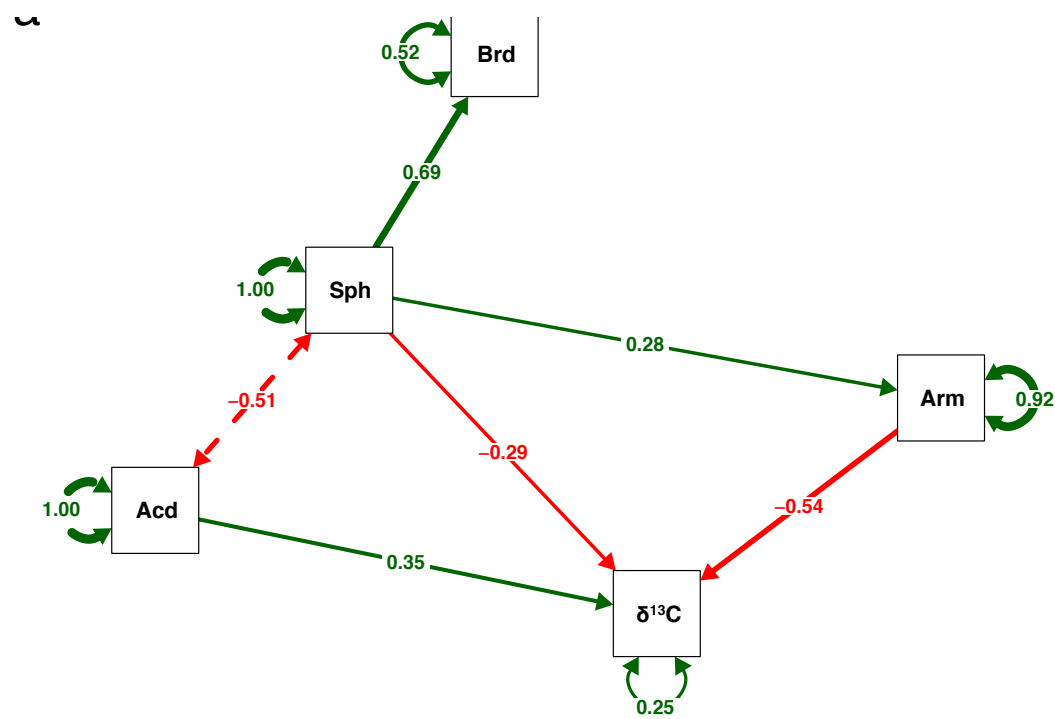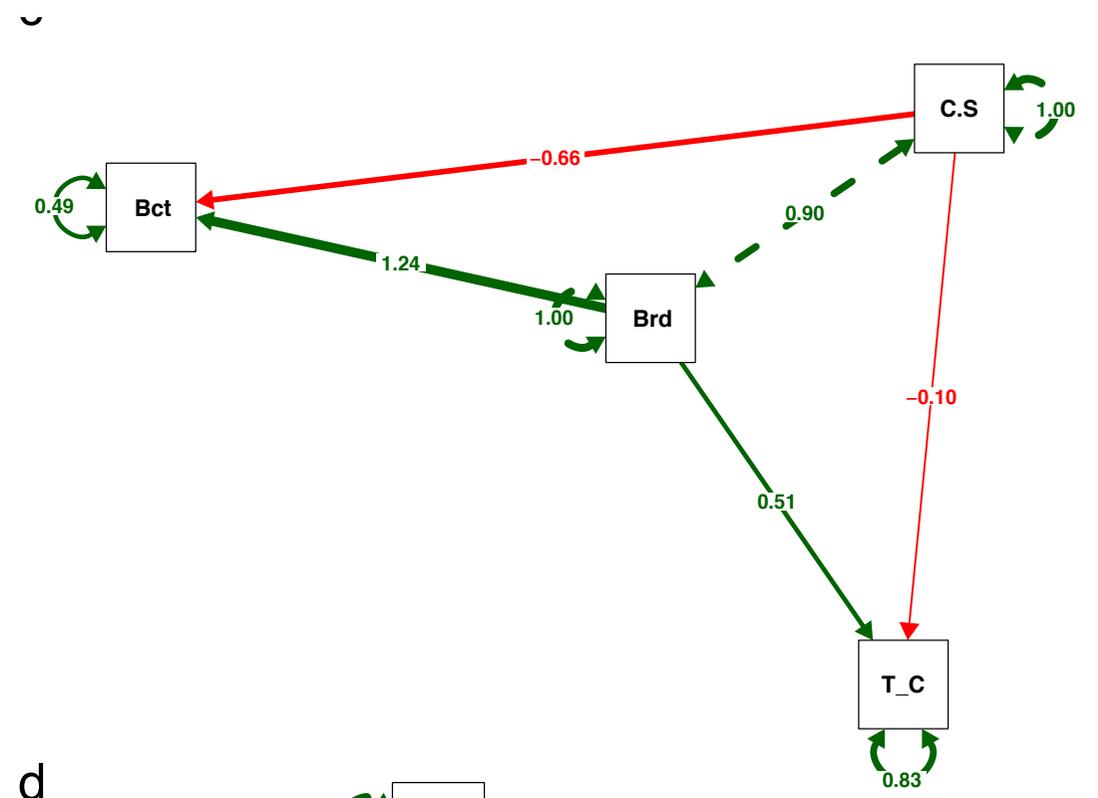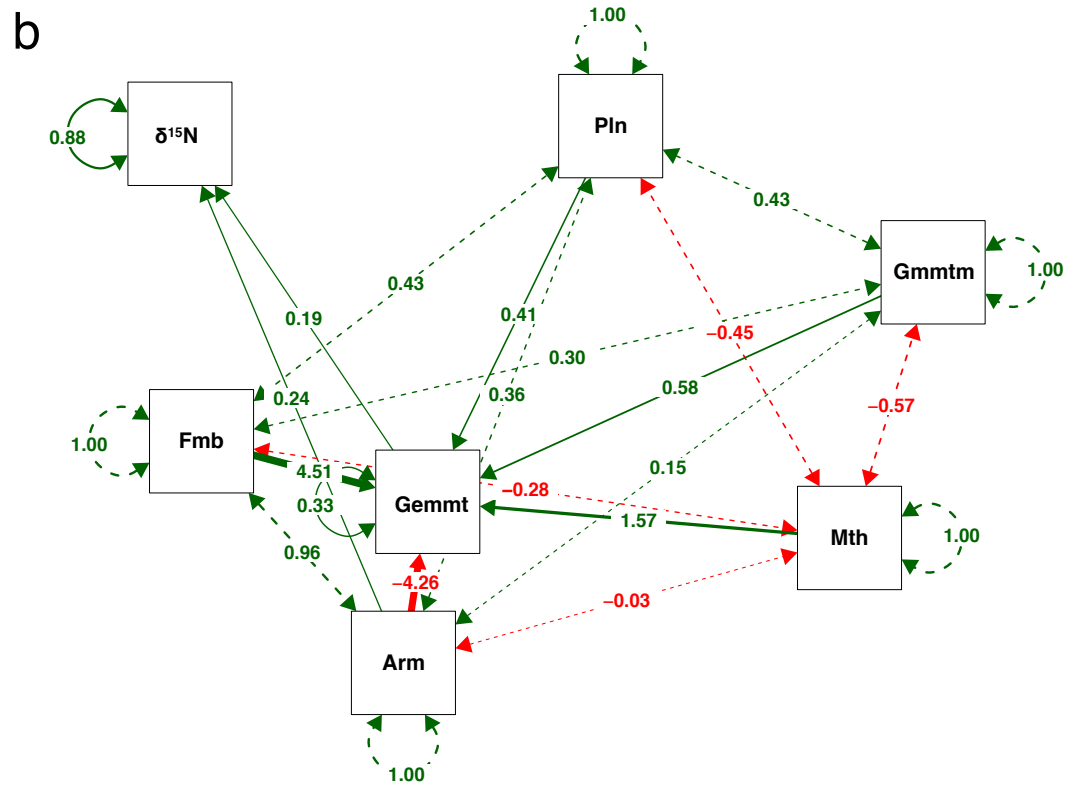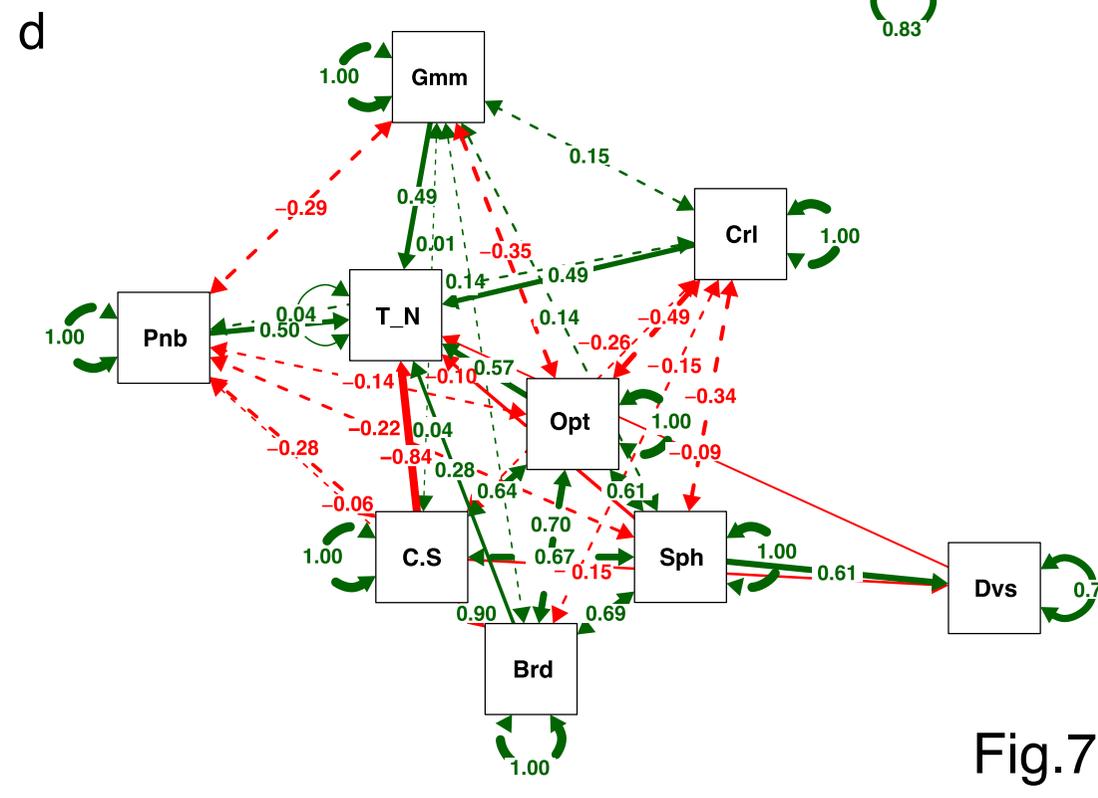

Fig.7

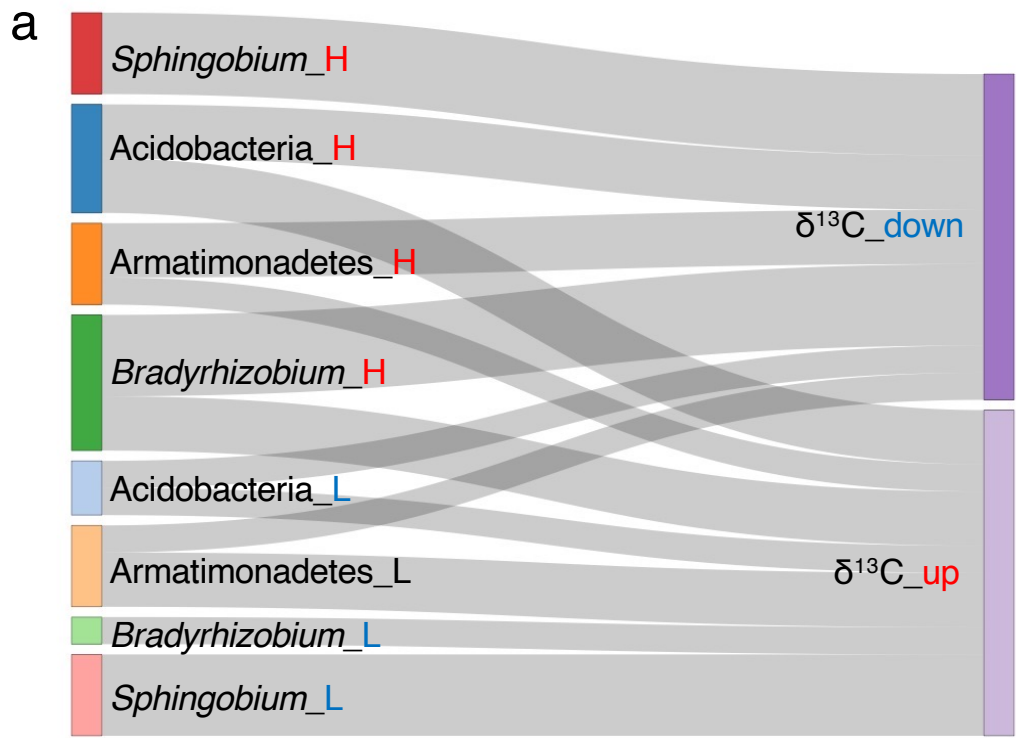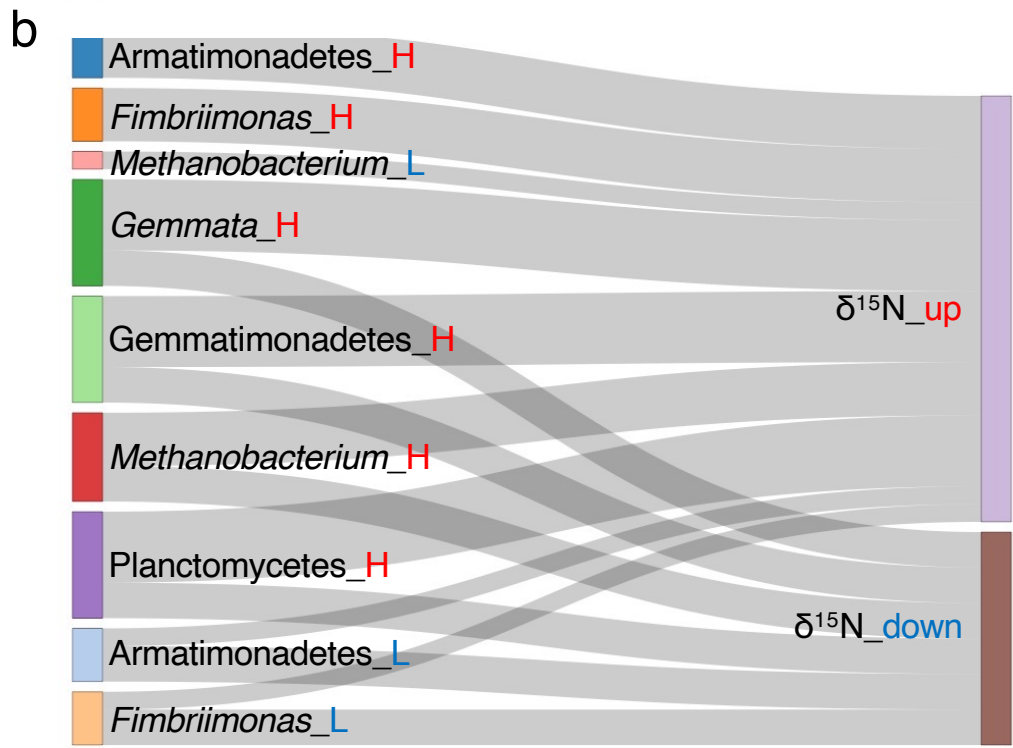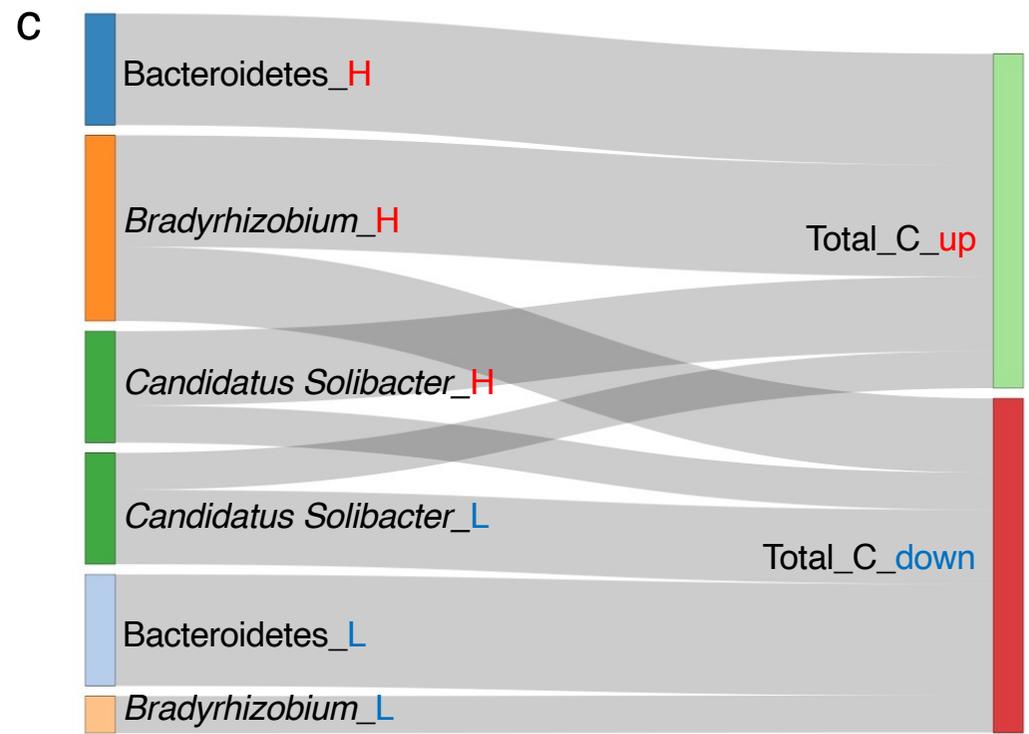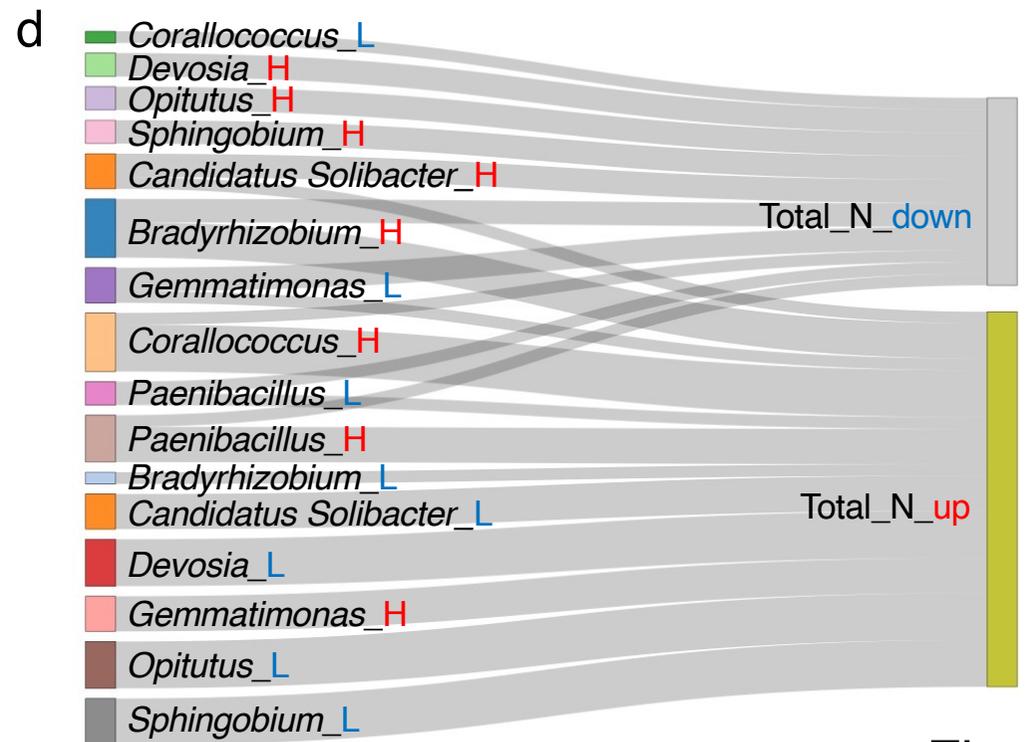

Fig.8

# Supplementary Information

# A potential network structure of symbiotic bacteria

# involved in carbon and nitrogen metabolism of wood-utilizing insect larvae.

Authors: Hirokuni Miyamoto*, Futo Asano, Koutarou Ishizawa, Wataru Suda, Hisashi Miyamoto, Naoko Tsuji[3], Makiko Matsuura, Arisa Tsuboi, Chitose Ishii, Teruno Nakaguma, Chie Shindo, Tamotsu Kato, Atsushi Kurotani, Hideaki Shima, Shigeharu Moriya, Masahira Hattori, Hiroaki Kodama, Hiroshi Ohno, Jun Kikuchi*

* Co-correspondence:

Hirokuni Miyamoto Ph.D., hirokuni.miyamoto@riken.jp, h-miyamoto@faculty.chiba-u.jp

Jun Kikuchi Ph.D., jun.kikuchi@riken.jp

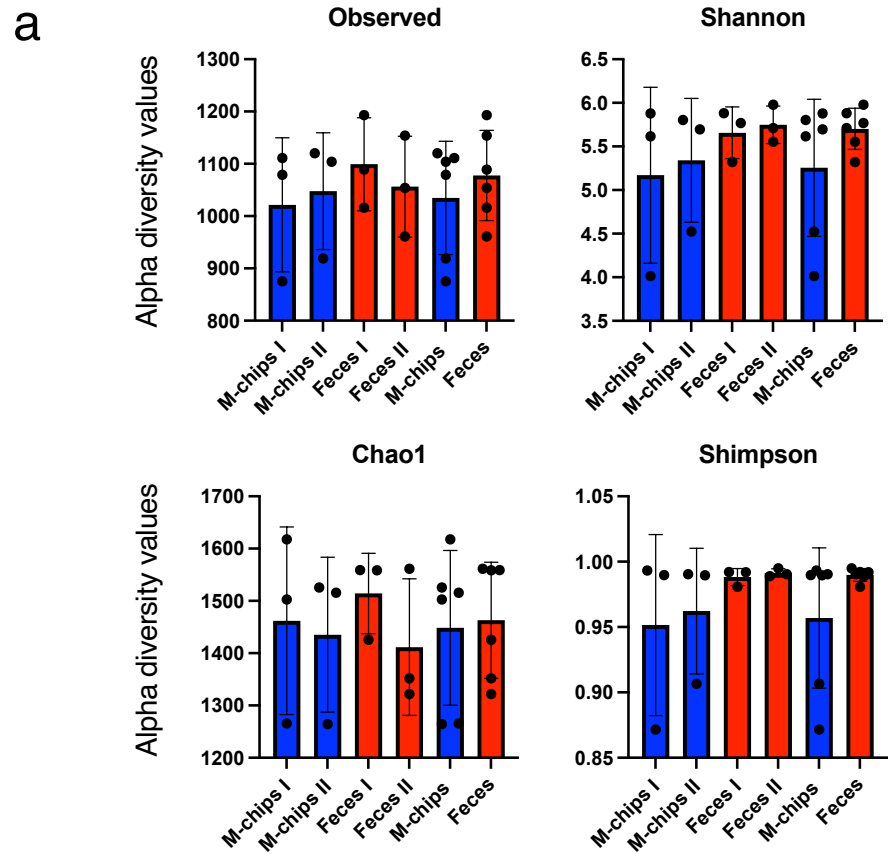
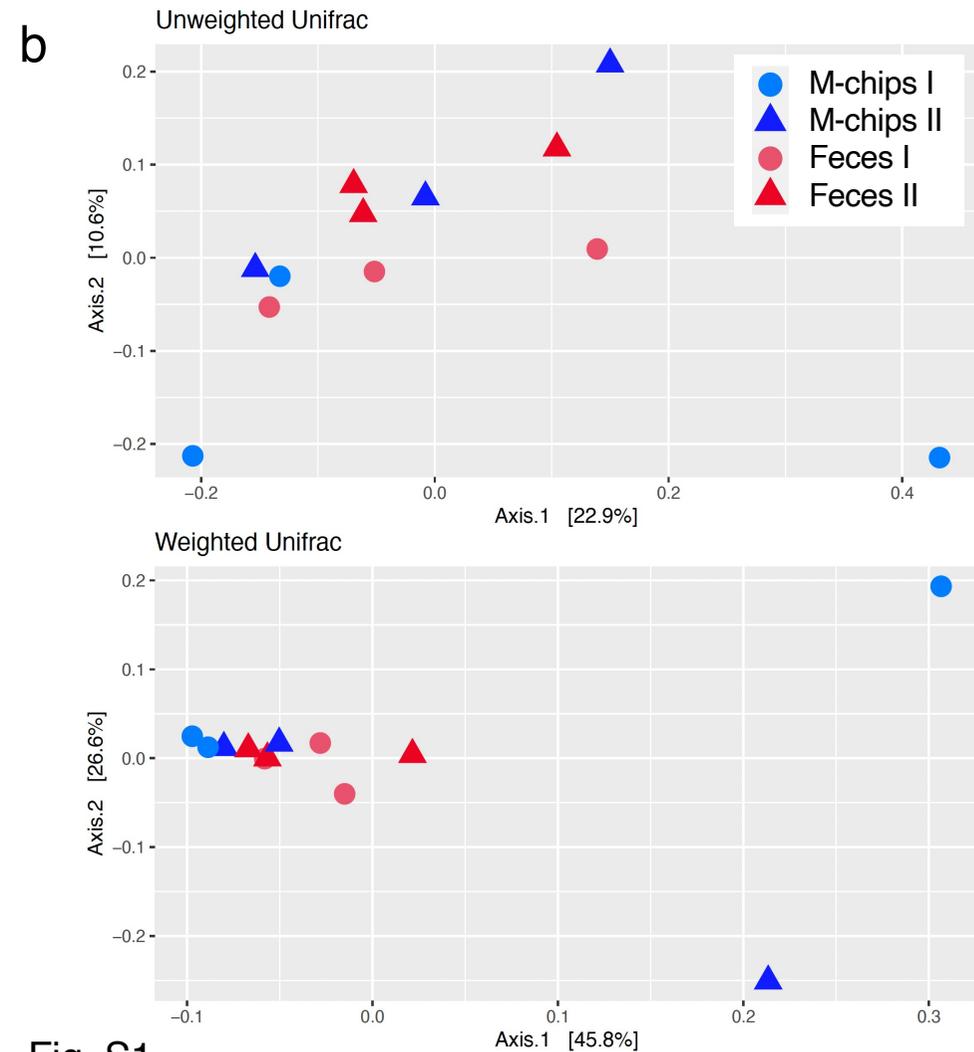

Fig. S1

OTU number and the Chao1, Shannon, and Simpson indices representing α-diversity in the habitat (M-chips) and feces (Feces). (b) UniFrac distance graph, unweighted and weighted, which shows β-diversities in the habitat (M-chips) and feces. (c) Values calculated based on unweighted and weighted data are shown under different environmental conditions. I and II show the environmental conditions: I, a group sprayed with water only; II, a group sprayed with 20% compost extract.

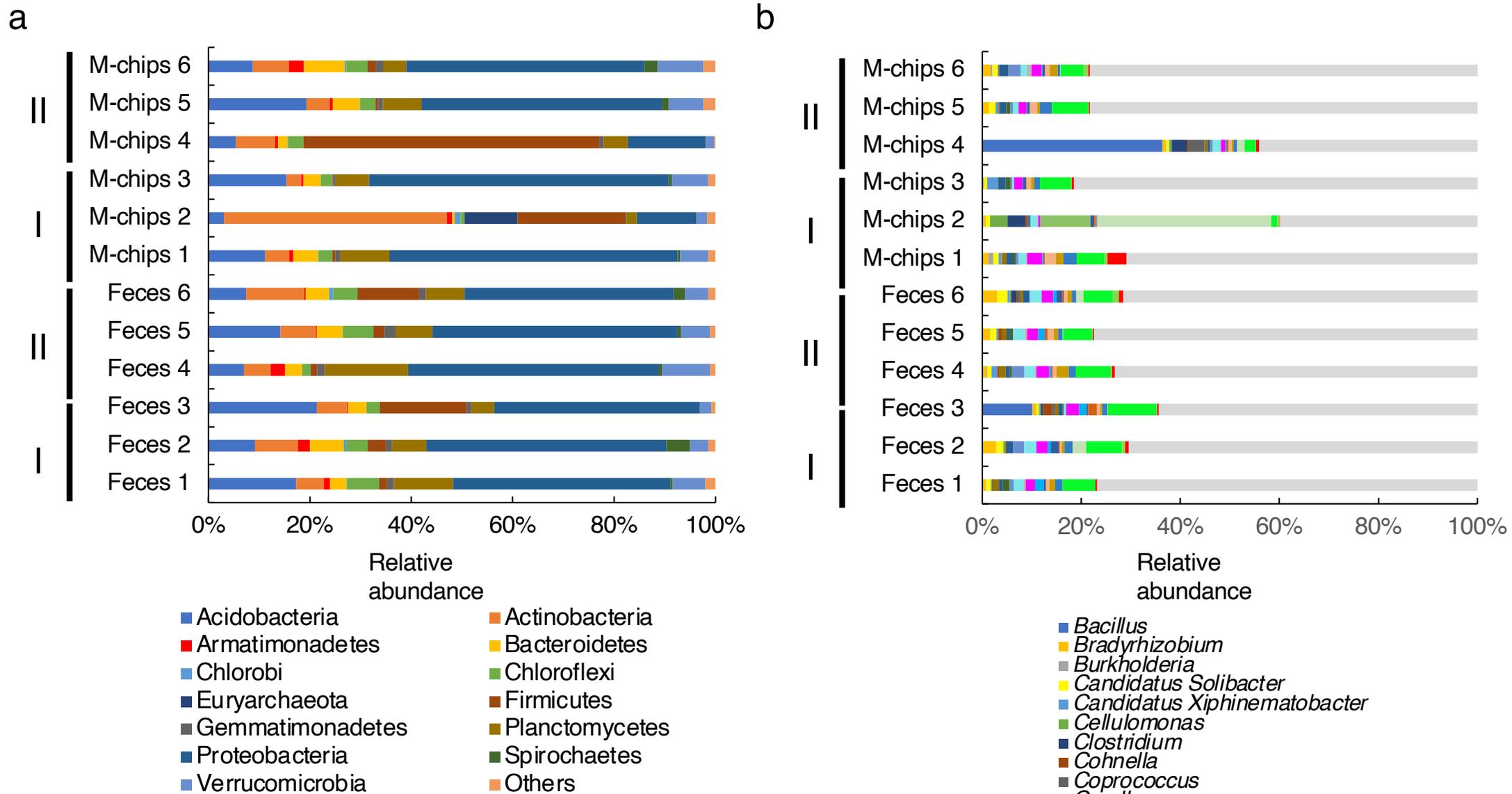

Fig. S2
Relative abundances of (a) phyla and (b) genera in the habitat (M-chips) and feces (Feces) under different environmental conditions. I and II show the environmental conditions: I, a group sprayed with water only; II, a group sprayed with 20% compost extract.

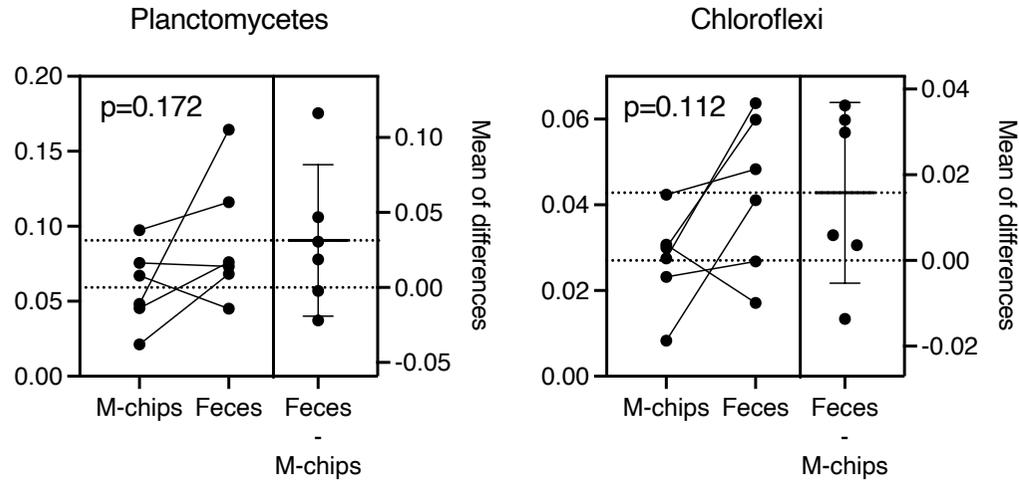
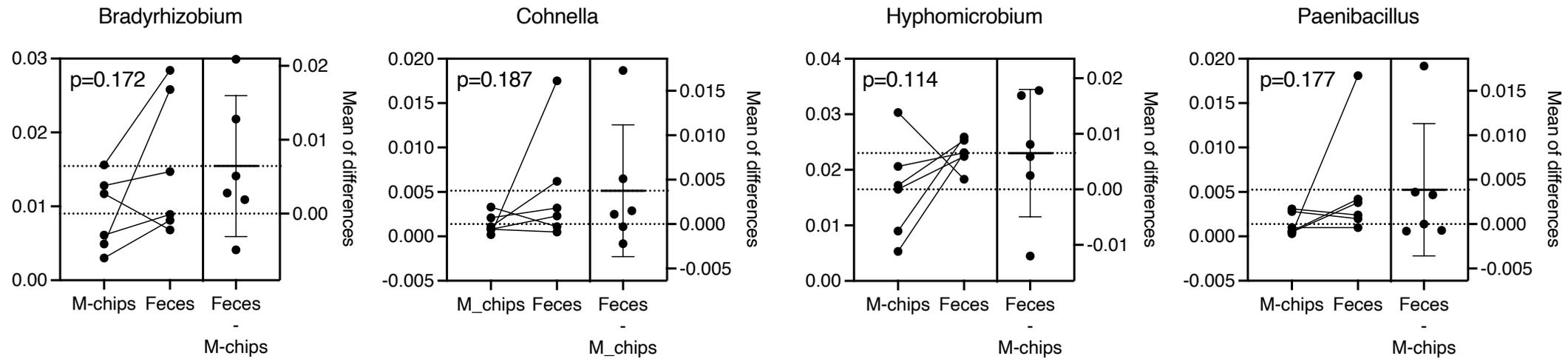

Fig. S3
Estimation plots of representative phyla and (b) genera in Fig. 2b with their significance values (0.1< p <0.2; >1% as the maximal value of the detected bacterial community among the whole community in each group).

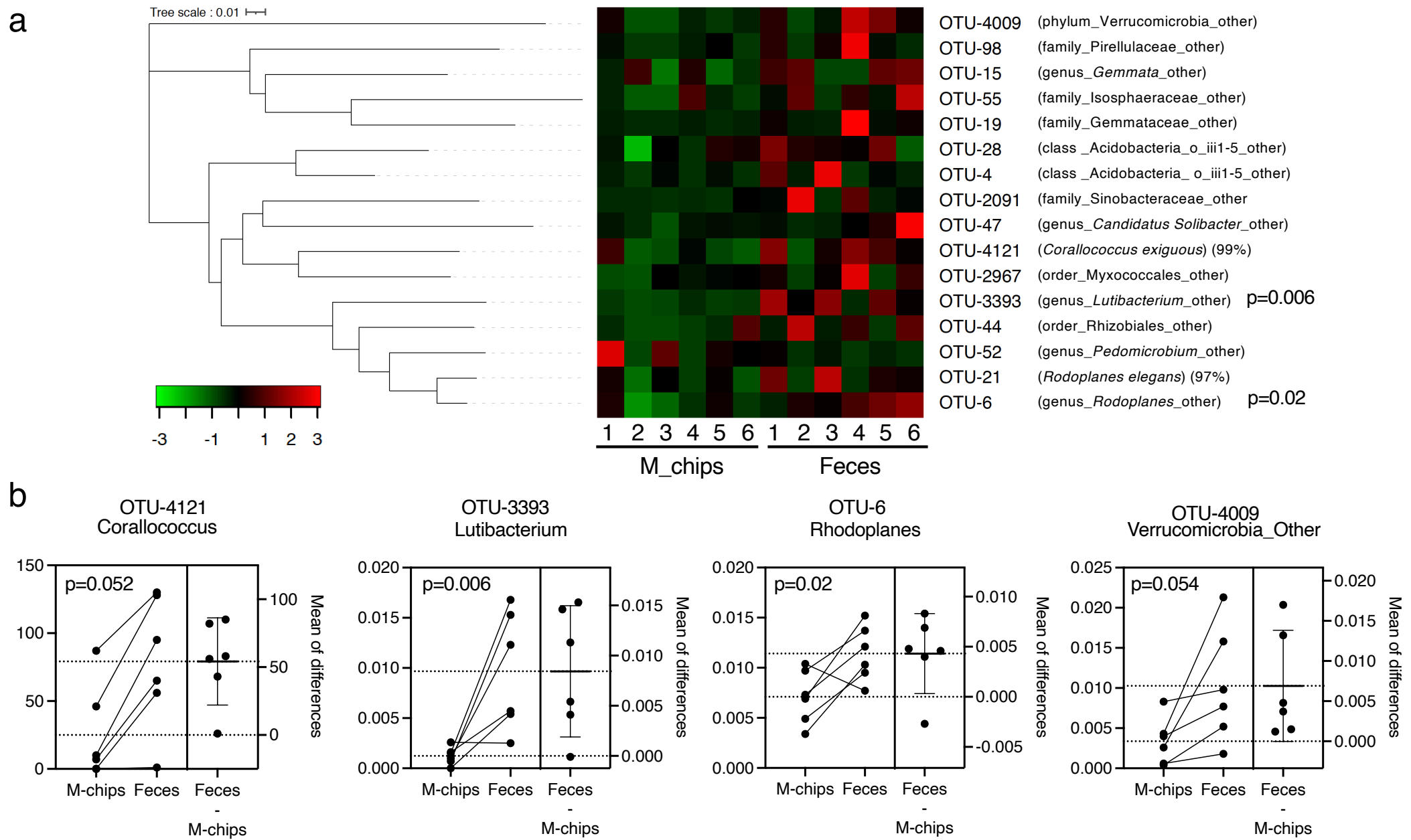

Fig. S4
Phylogenetic tree and heatmap of representative bacterial OTUs in the feces (p<0.2; >1% as the maximal value of the detected bacterial community among the whole community in each group).
Estimation plots of representative OTUs in Fig. 4a with their significance values (p<0.1; >1% as the maximal value of the detected bacterial communities among the whole community in each group).

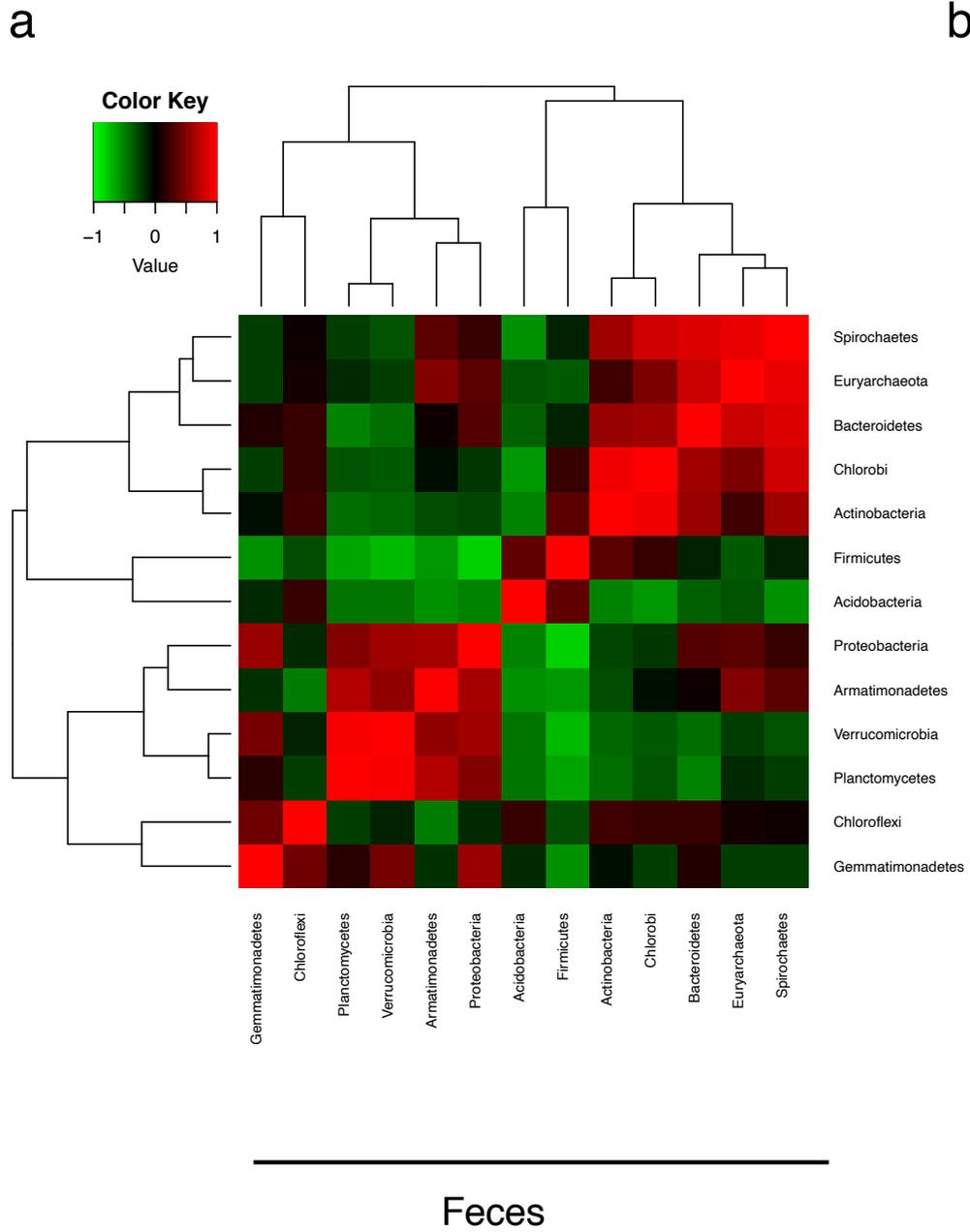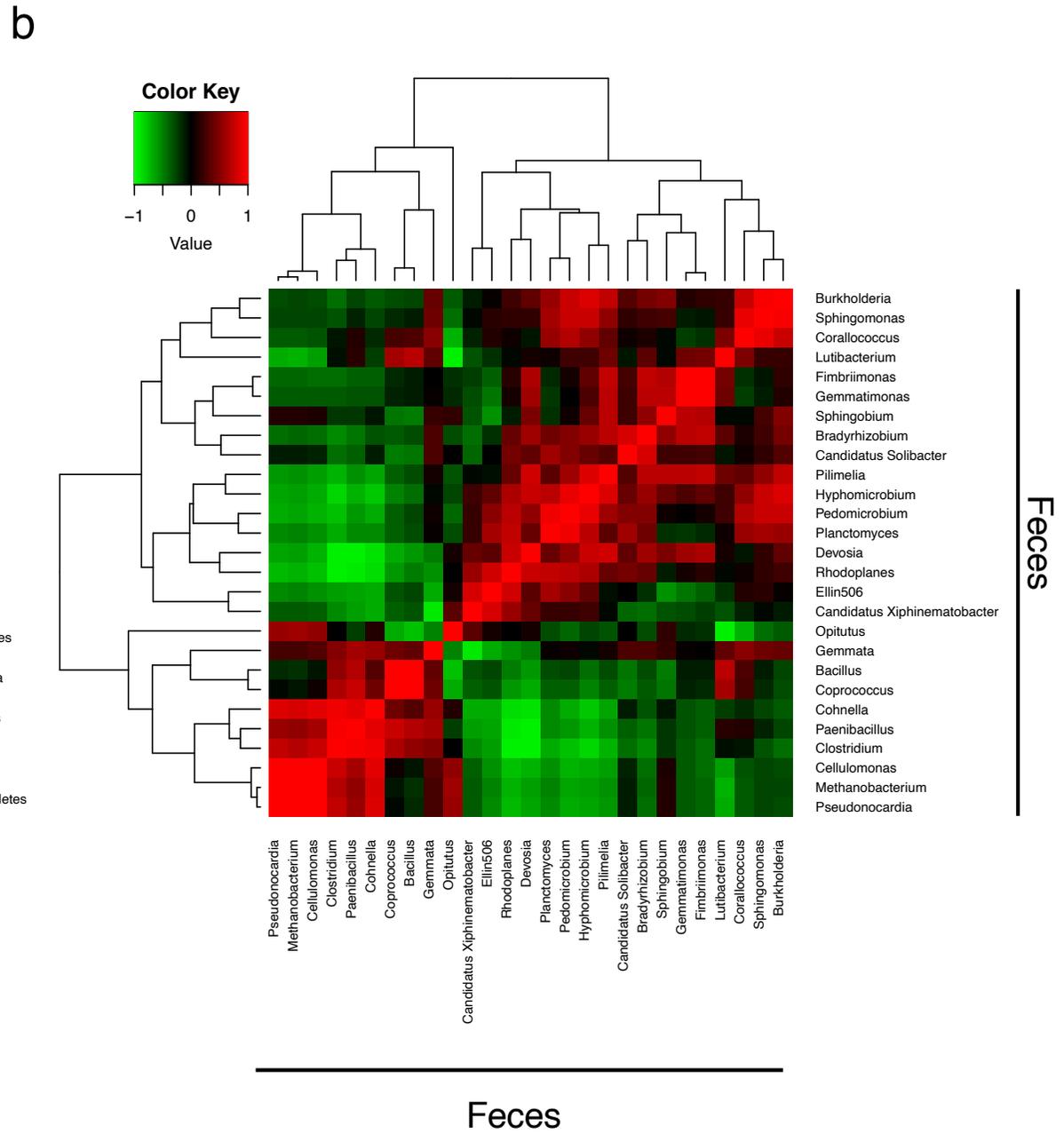

Fig. S5
Correlation heatmaps of the bacterial community in the feces of beetle larvae: (a) phyla and (b) genera.

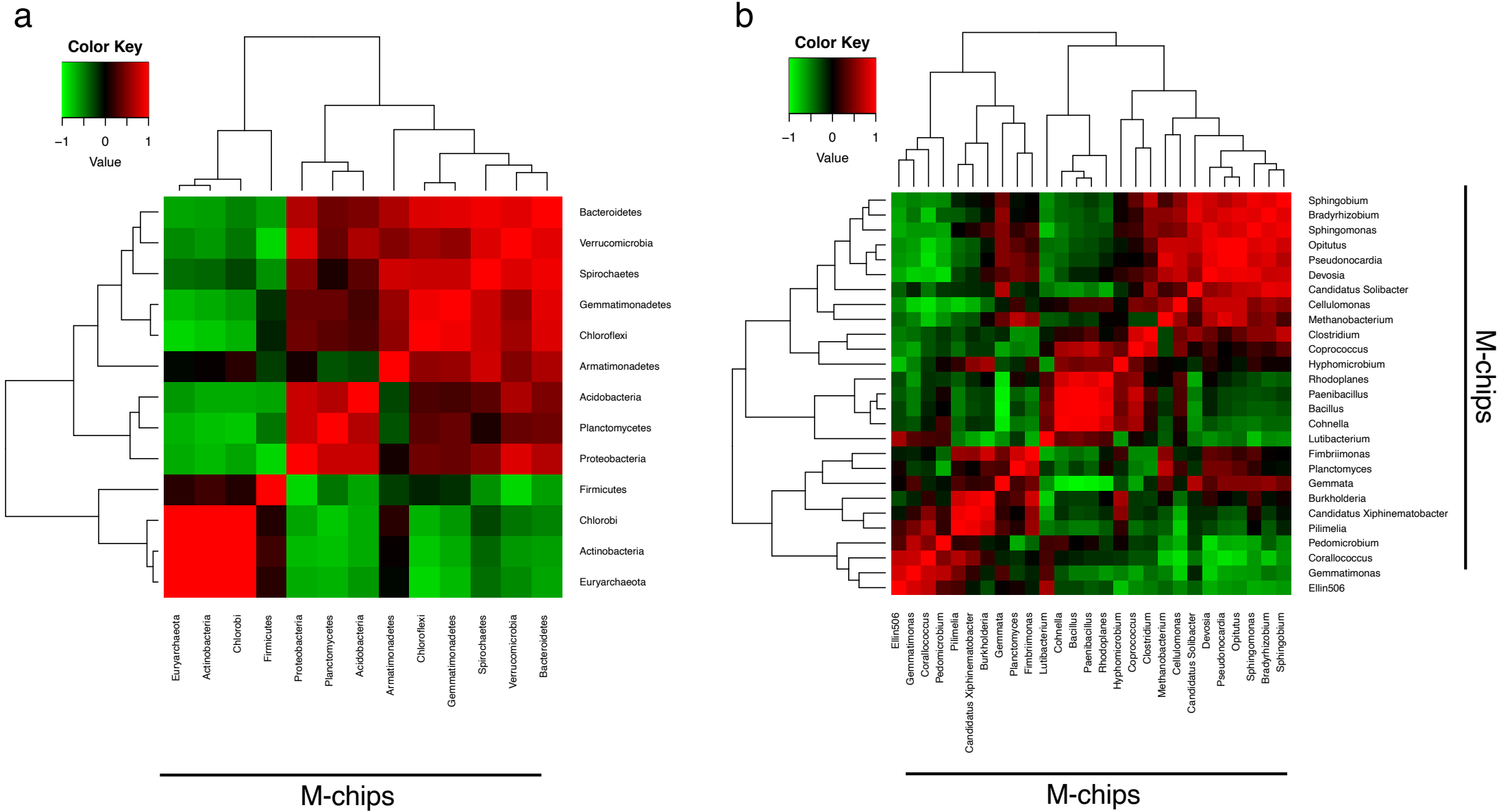

Fig. S6
Correlation heatmaps of the bacterial community in the habitat of beetle larvae alone (a) phyla and (b) genera.

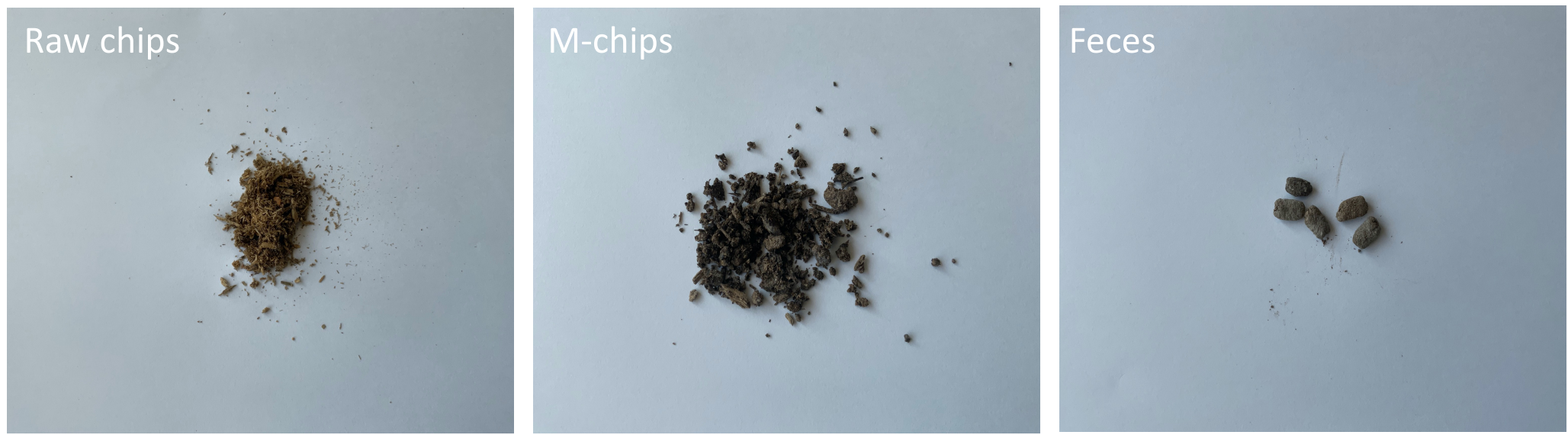

| Component | Raw-chips (3) | M-chips | | | Feces | | |
|---|---|---|---|---|---|---|---|
| | | Condition I (3) | Condition II (3) | I + II (6) | I (3) | II (3) | I + II (6) |
| $\delta^{13}C$ | -27.39 ± 0.01 | -27.85 ± 0.04 | -27.82 ± 0.15 | -27.85 ± 0.04 [b] | -27.63 ± 0.15 | -27.86 ± 0.12 | -27.75 ± 0.10 [c] |
| $\delta^{15}N$ | -2.92 ± 0.08 | 1.45 ± 0.06 | 0.77 ± 0.07 [a] | 1.11 ± 0.16 [b] | 1.83 ± 0.22 | 1.33 ± 0.61 | 1.58 ± 0.31 [c] |
| Total C (%) | 47.81 ± 0.10 | 40.68 ± 2.90 | 44.01 ± 1.76 | 42.34 ± 1.69 [b] | 42.48 ± 0.85 | 43.69 ± 1.14 | 43.08 ± 0.69 [c] |
| Total N (%) | 0.34 ± 0.01 | 1.16 ± 0.10 | 1.24 ± 0.27 | 1.20 ± 0.13 [b] | 1.93 ± 0.34 [d] | 1.44 ± 0.27 | 1.69 ± 0.22 [cd] |
| C/N ratio | 142.5 ± 2.34 | 35.48 ± 2.97 | 40.26 ± 11.3 | 37.87 ± 5.35 [b] | 23.82 ± 5.24 | 33.34 ± 8.5 | 28.59 ± 4.95 [c] |

Fig. S7
Photographs of the samples used in this experiment and (b) stable isotope ($\delta^{13}C$ and $\delta^{15}N$) levels, carbon and nitrogen levels, and carbon/nitrogen ratios in the fresh wood chips of the habitat (raw chips), decayed chips (M-chips), and larval feces (Feces). I and II show the environmental conditions: I, a group sprayed with water only; II, a group sprayed with 20% compost extract.

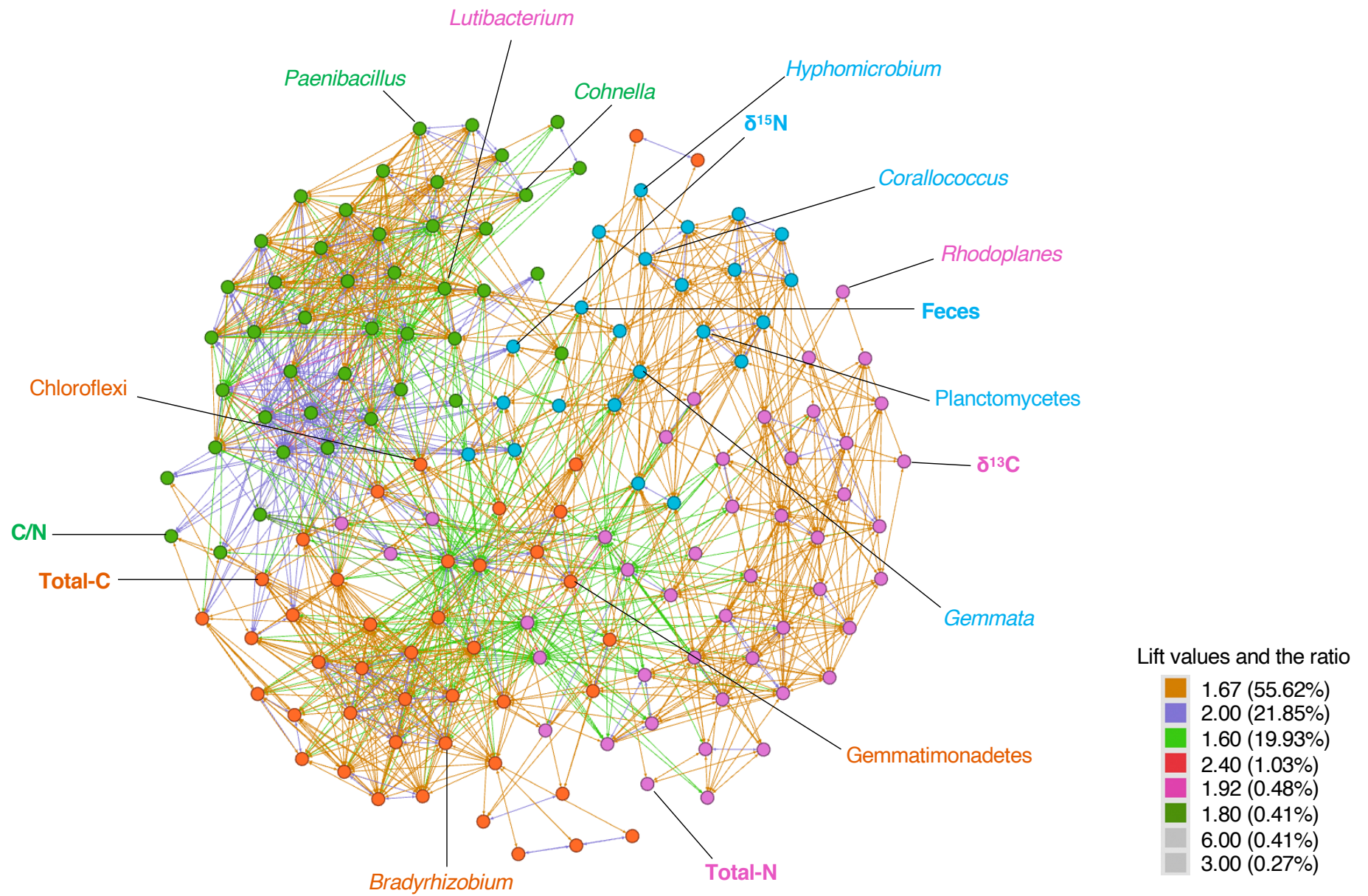

Fig. S8
Systemic networks of factors associated with chemical indices and feces. The factors shown in Figs. 2bc and 4a are indicated with black lines. Modularity classes are discriminated by four colors and lift values, and the ratio are shown.

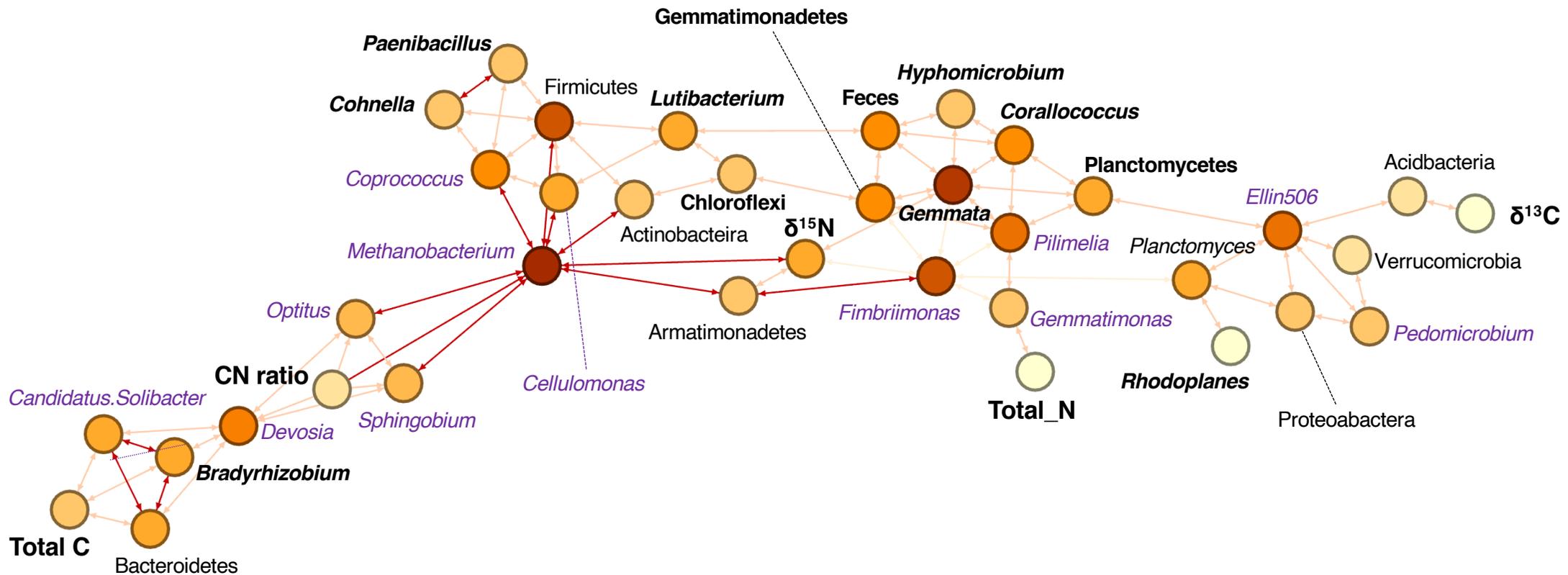

Fig. S9
Systemic networks of factors shown in Figs. 2bc and 4a. The difference in the color of the nodes indicates degree strength, the value of which is the sum of the weights of the adjacent edges for each node. The bacteria and components in Figs. 2b and 6a are shown in bold letters. Stable bacteria, *i.e.*, those that were not significantly different between the habitat and the feces, are shown in violet.

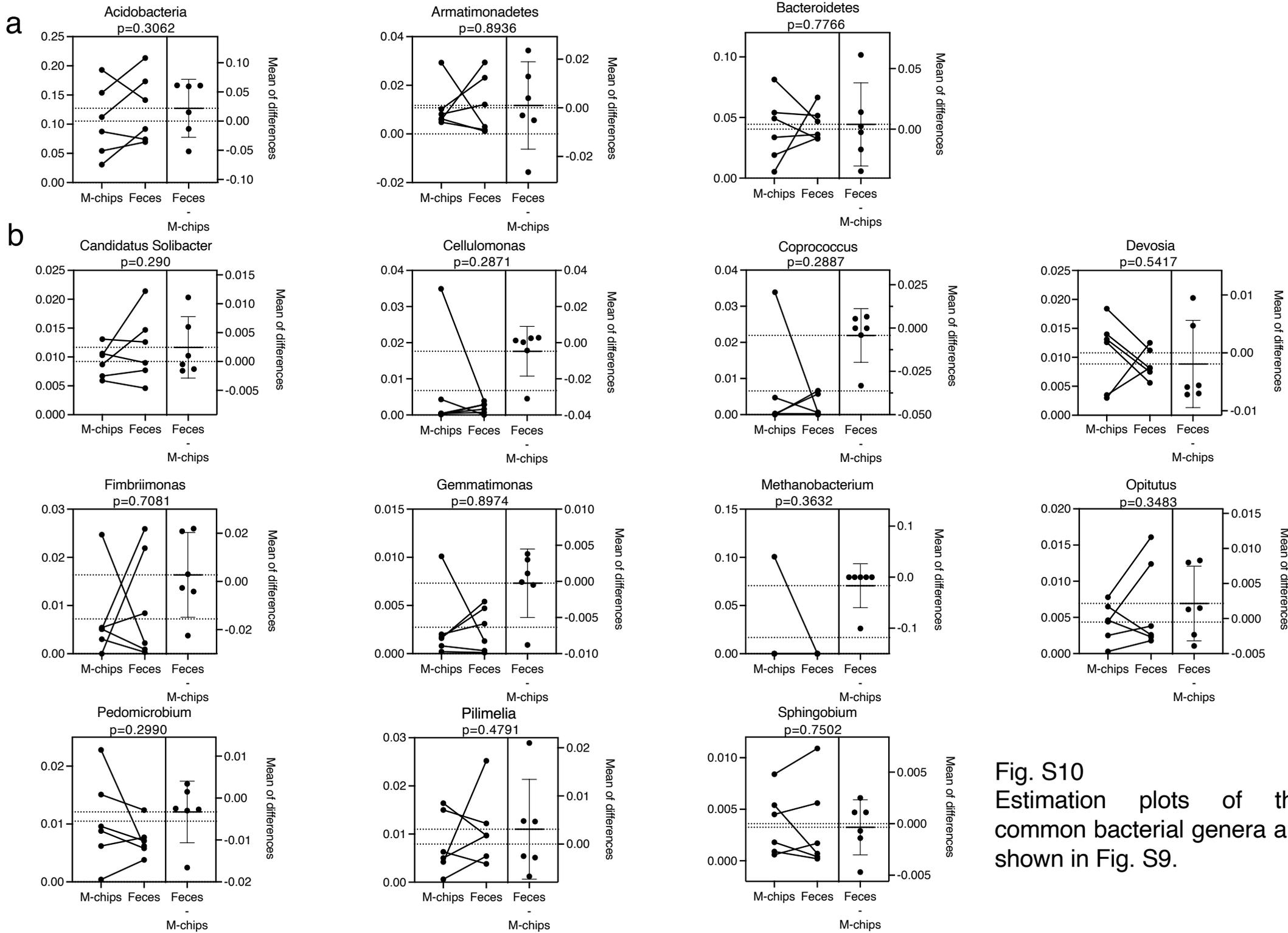

Fig. S10 Estimation plots of the common bacterial genera are shown in Fig. S9.

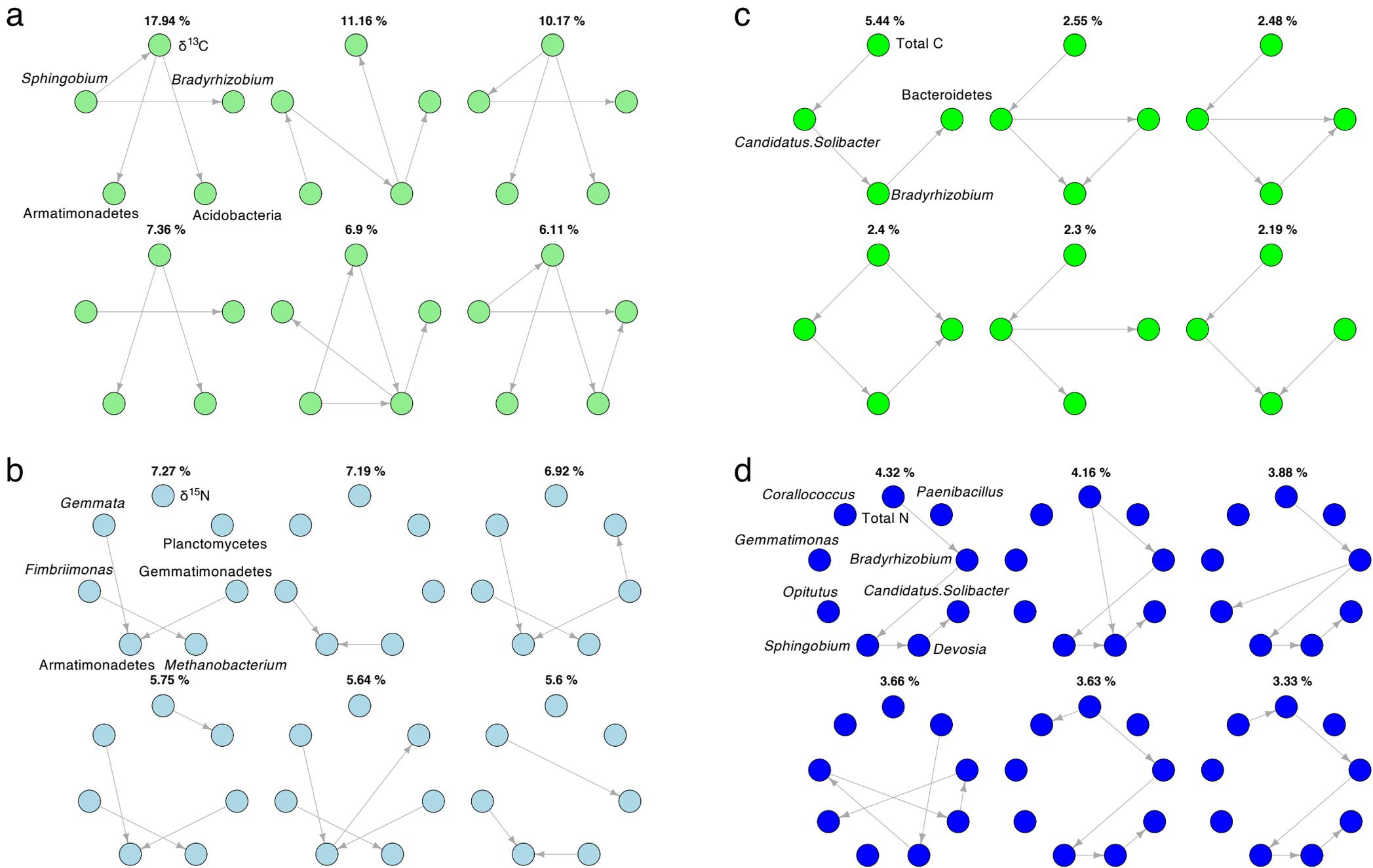

Fig. S11
The top six groups were identified by structural equation modeling and estimated by BayesLiNGAM visualized with percentages. Finally, the most appropriate groups calculated by structural equation modeling based on mediation values (M-chips and feces) in Fig. 7 were selected.

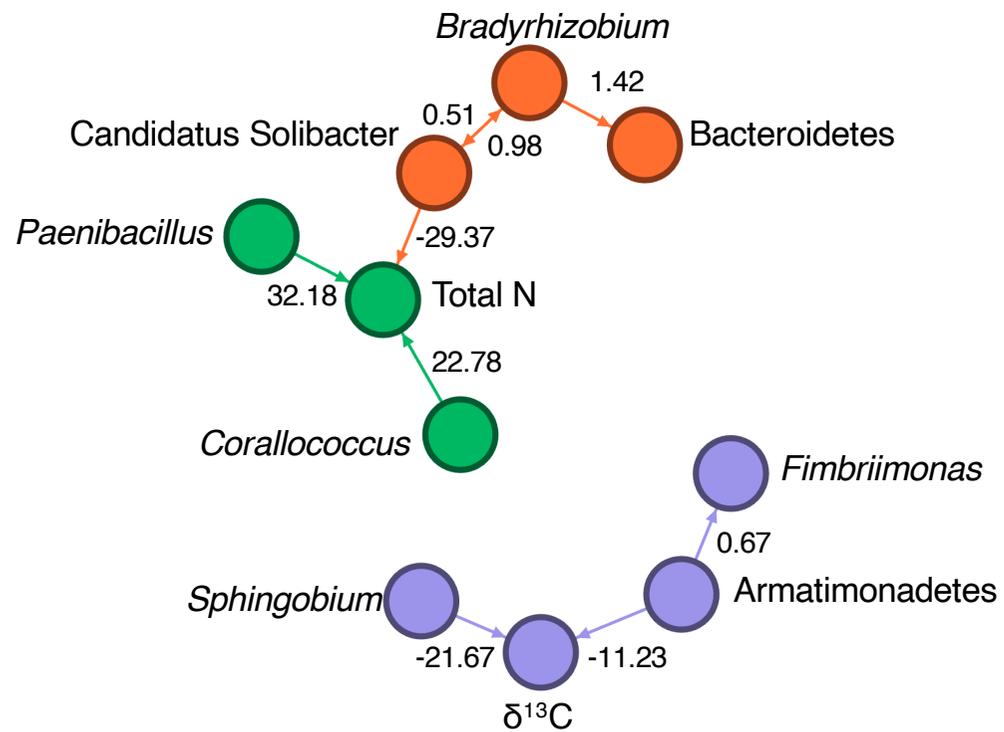

Fig. S12
Causal relationship estimated by the LiNGAM. The LiNGAM values show the extent of contribution.

Table S1.
Body weights of larvae were used in this experiment. One month, one male larva and one female larva were bred in the same box. I and II show the environmental conditions: I, a group sprayed with water only; II, a group sprayed with 20% compost extract.

| Condition | Category | | Body weight of larvae (g) | | | |
| --- | --- | --- | --- | --- | --- | --- |
| | | | 2018/5/6 | 2018/5/20 | 2018/5/27 | 2018/6/9 |
| I | Box 1 | male | 22 | 22 | 24 | 21 |
| | | female | 19 | 19 | 18 | 16 |
| | Box 2 | male | 23 | 23 | 17 | 27 |
| | | female | 19 | 19 | 22 | 19 |
| | Box 3 | male | 21 | 21 | 20 | 18 |
| | | female | 17 | 17 | 17 | 17 |
| | Box 4 | male | 20 | 20 | 20 | 20 |
| | | female | 17 | 17 | 18 | 19 |
| II | Box 5 | male | 25 | 25 | 27 | 17 |
| | | female | 14 | 14 | 14 | 11 |
| | Box 6 | male | 22 | 22 | 22 | 17 |
| | | female | 17 | 17 | 18 | 17 |

↑
plus New chips

Table S2.
Statistical values of the final optimal structural equation models for $\delta^{13}C$, $\delta^{15}N$, total carbon, and total nitrogen shown in Fig. 7. Abbreviations in the table mean the following: Delta13C, $\delta^{13}C$; Delta15N, $\delta^{15}N$; Total_C, Total carbon; Total_N, Total nitrogen; chisq, chi-square $\chi^2$; df, degrees of freedom; p-value, p values (chi-square); cfi, comparative fix index; tli, Tucker–Lewis index; nfi, normed fit index; rfi, relative fit index; SRMR, standardized root mean residuals; AIC, Akaike information criterion; rmsea, root mean square error of approximation; gfi, goodness-of-fit index; agfi, adjusted goodness-of-fit index. Column No. 1 shows the best numerical structural equation model for $\delta^{13}C$, $\delta^{15}N$, total carbon, and total nitrogen. Column No. 2 shows the inferior numerical structural equation models.

| Category | | No.1 | | | No.2 | | |
|---|---|---|---|---|---|---|---|
| $\delta^{13}C$ | Model | **Delta13C ~ Sphingobium + Armatimonadetes + Acidobacteria** | | | **Delta13C ~ Sphingobium + Armatimonadetes** | | |
| | | **Bradyrhizobium + Armatimonadetes ~ Sphingobium** | | | **Armatimonadetes ~ Sphingobium** | | |
| | Fit indices | chisq 2.364 | df 4 | pvalue 0.669 | chisq 1.220 | df 1 | pvalue 0.269 |
| | | cfi 1.000 | tli 1.186 | rfi 0.815 | cfi 985 | tli 0.926 | rfi 0.693 |
| | | nfi 0.917 | SRMR 0.0986 | AIC -175.37 | nfi 0.939 | SRMR 0.041 | AIC -88.580 |
| | | rmsea 0 | gfi 0.953 | agfi 0.823 | rmsea 0.136 | gfi 0.950 | agf 0.499 |
| $\delta^{15}N$ | Model | **Delta15N ~ Gemmata** | | | **Delta15N ~ Gemmata** | | |
| | | **Gemmata ~ Fimbriimonas + Armatimonadetes + Methanobacterium** | | | **Gemmata ~ Fimbriimonas + Armatimonadetes + Methanobacterium** | | |
| | | **Gemmata ~ Gemmatimonadetes + Planctomycetes** | | | **Gemmata ~ Gemmatimonadetes** | | |
| | | **Delta15N ~ Armatimonadetes** | | | | | |
| | Fit indices | chisq 2.456 | df 4 | pvalue 0.652 | chisq 1.616 | df 4 | pvalue 0.806 |
| | | cfi 1.000 | tli 1.670 | rfi 0.610 | cfi 1.000 | tli 2.582 | rfi 0.707 |
| | | nfi 0.858 | SRMR 0.047 | AIC -61.05 | nfi 0.870 | SRMR 0.065 | AIC -60.991 |
| | | rmsea 0 | gfi 0.944 | agfi 0.611 | rmsea 0 | gfi 0.936 | agfi 0.664 |
| Total C | Model | **Total_C ~ Candidatus.Solibacter + Bradyrhizobium** | | | **Total_C ~ Candidatus.Solibacter + Bradyrhizobium** | | |
| | | **Bacteroidetes ~ Candidatus.Solibacter + Bradyrhizobium** | | | **Bradyrhizobium ~ Devosia** | | |
| | | | | | **Candidatus.Solibacter ~ Devosia** | | |
| | Fit indices | chisq 0.051 | df 1 | pvalue 0.822 | chisq 19.289 | df 2 | pvalue 6.48E-05 |
| | | cfi 1 | tli 1.533 | rfi 0.982 | cfi 0.051 | tli -1.848 | rfi 1 |
| | | nfi 0.996 | SRMR 0.013 | AIC 0.640 | nfi 0.203 | SRMR 0.287 | AIC -110.42 |
| | | rmsea 0 | gfi 0.998 | agfi 0.975 | rmsea 0.849 | gfi 0.625 | agfi -0.876 |
| Total N | Model | **Total_N ~ Corallococcus + Gemmatimonas + Opitutus + Sphingobium** | | | **Total_N ~ Corallococcus + Gemmatimonas + Opitutus + Devosia** | | |
| | | **Total_N ~ Devosia + Candidatus.Solibacter + Bradyrhizobium + Paenibacillus** | | | **Total_N ~ Candidatus.Solibacter + Bradyrhizobium + Paenibacillus + Cohnella** | | |
| | | **Devosia ~ Candidatus.Solibacter + Sphingobium** | | | **Devosia ~ Candidatus.Solibacter** | | |
| | Fit indices | chisq 4.445 | df 5 | pvalue 0.487 | chisq 6.874 | df 6 | pvalue 0.333 |
| | | cfi 1 | tli 1.051 | rfi 0.720 | cfi 0.967 | tli 0.918 | rfi 0.589 |
| | | nfi 0.907 | SRMR 0.055 | AIC -99.03 | nfi 0.836 | SRMR 0.090 | AIC -92.706 |
| | | rmsea 0 | gfi 0.999 | agfi 0.991 | rmsea 0.110 | gfi 0.996 | agfi 0.971 |

Table S3.
A list of models targeted by causal mediation analysis (CMA) for δ¹³C in Fig. 7 and their statistical values. Abbreviations in the table mean the following : Delta13C, δ¹³C; ACME, average causal mediation effects; ADE, average direct effect; *, p <0.05; **, p<0.01; ***, p<0.001.

| | | | | | | |
|---|---|---|---|---|---|---|
| | | δ¹³C regression models | | | | (I)Delta13C ~ Sphingobium + Armatimonadetes + Acidobacteria |
| | | | | | | (II)Bradyrhizobium + Armatimonadetes ~ Sphingobium |
| (I) | Estimate std. | Error | t value | Pr ( > ltl) | | Nonparametric bootstrap Confidence Intervals |
| (Intercept) | -27.7588 | 0.1271 | -218.319 | <2e-16 | *** | |
| Sphingobium | -16.003 | 11.3858 | -1.406 | 0.1975 | | - |
| Armatimonadetes | -10.259 | 3.6756 | -2.791 | 0.0235 | * | |
| Acidobacteria | 1.1716 | 0.7238 | 1.619 | 0.1442 | | |
| | | | | | | Quasi-Bayesian Confidence Intervals |
| (II) | Estimate std. | Error | t value | Pr ( > ltl) | | |
| (Intercept) | 0.015315 | 0.004791 | 3.197 | 0.00954 | ** | - |
| Sphingobium | 2.377669 | 0.995071 | 2.389 | 0.03799 | * | |

Table S4.
A list of models targeted by causal mediation analysis for δ¹⁵N in Fig. 7 and their statistical values. Abbreviations in the table mean the following: Delta15N, $\delta^{15}N$; ACME, average causal mediation effect; ADE, average direct effect; #, $p < 0.1$.

| δ¹⁵N regression models | | | | | | (I) Delta15N ~ Gemmata + Armatimonadetes | | | | |
| --- | --- | --- | --- | --- | --- | --- | --- | --- | --- | --- |
| | | | | | | (II) Gemmata ~ Fimbriimonas + Armatimonadetes + Methanobacterium + Gemmatimonadetes + Planctomycetes | | | | |

| | | | | | | Nonparametric bootstrap Confidence Intervals | | | | |
| --- | --- | --- | --- | --- | --- | --- | --- | --- | --- | --- |
| (I) | Estimate std. | Error | t value | Pr ( > |t|) | | | Estimate 95% | CI lower 95% | CI upper | p-value |
| (Intercept) | 0.9237 | 0.4828 | 1.913 | 0.088 | # | ACME | 0 | 0 | 0 | 1 |
| Gemmata | 16.8272 | 28.8199 | 0.584 | 0.574 | | ADE | -2.92 | -10.71 | 8.02 | 0.48 |
| Armatimonadetes | 14.2315 | 19.7173 | 0.722 | 0.489 | | Total Effect | -2.92 | -10.71 | 8.02 | 0.48 |
| | | | | | | Prop.Mediated | 0 | 0 | 0 | 1 |
| | | | | | | Sample Size Used :12  Simulations: 1000 | | | | |
| | | | | | | Quasi-Bayesian Confidence Intervals | | | | |
| (II) | Estimate std. | Error | t value | Pr ( > |t|) | | | Estimate 95% | CI lower 95% | CI upper | p-value |
| (Intercept) | 0.003248 | 0.005729 | 0.567 | 0.5913 | | ACME | 0 | 0 | 0 | 1 |
| Fimbriimonas | 3.261639 | 1.780156 | 1.832 | 0.1166 | | ADE | -2.91 | -6.17 | 0.1 | 0.066 |
| Armatimonadetes | -2.915195 | 1.63036 | -1.788 | 0.124 | | Total Effect | -2.91 | -6.17 | 0.1 | 0.066 |
| Methanobacterium | 0.380351 | 0.164101 | 2.318 | 0.0596 | # | Prop.Mediated | 0 | 0 | 0 | 1 |
| Gemmatimonadetes | 0.746686 | 0.387198 | 1.928 | 0.1021 | | | | | | |
| Planctomycetes | 0.077369 | 0.05535 | 1.398 | 0.2117 | | Sample Size Used :12  Simulations: 1000 | | | | |

Table S5.
A list of models targeted by causal mediation analysis for total carbon in Fig. 7 and their statistical values. Abbreviations in the table mean the the following: Total_C, Total carbon; ACME, average causal mediation effect; ADE, average direct effect; *, p <0.05; **, p<0.01; ***, p<0.001.

| Total C regression models | | | | | | (I)Total_C ~ Candidatus.Solibacter + Bradyrhizobium | | | | | |
| --- | --- | --- | --- | --- | --- | --- | --- | --- | --- | --- | --- |
| | | | | | | (II)Bradyrhizobium + Armatimonadetes ~ Sphingobium | | | | | |
| | | | | | | Nonparametric bootstrap Confidence Intervals | | | | | |
| (I) | Estimate std. | Error | t value | Pr ( > |t|) | | | Estimate 95% | CI lower 95% | CI upper | p-value | |
| (Intercept) | 41.06 | 2.503 | 16.405 | 5.17E-08 | *** | ACME | -216.506 | -7347.545 | 2595.02 | 0.812 | |
| Candidatus.Solibacter | -68.061 | 451.982 | -0.151 | 0.884 | | ADE | -2.938 | -8.985 | 1.92 | 0.286 | |
| Bradyrhizobium | 193.52 | 260.259 | 0.744 | 0.476 | | Total Effect | -219.443 | -7348.653 | 2589.88 | 0.814 | |
| | | | | | | Prop.Mediated | 0.987 | 0.965 | 1.03 | 0.002 | ** |
| | | | | | | Sample Size Used :12 | | Simulations: 1000 | | | |
| | | | | | | Quasi-Bayesian Confidence Intervals | | | | | |
| (II) | Estimate std. | Error | t value | Pr ( > |t|) | | | Estimate 95% | CI lower 95% | CI upper | p-value | |
| (Intercept) | 0.03413 | 0.01305 | 2.616 | 0.028 | * | ACME | -252.719 | -3544.263 | 2984.88 | 0.89 | |
| Candidatus.Solibacter | -2.93776 | 2.35573 | -1.247 | 0.2438 | | ADE | -2.908 | -7.825 | 1.79 | 0.212 | |
| Bradyrhizobium | 3.18105 | 1.35647 | 2.345 | 0.0437 | * | Total Effect | -255.627 | -3547.675 | 2981.96 | 0.89 | |
| | | | | | | Prop.Mediated | 0.999 | 0.967 | 1.04 | 0.004 | ** |
| | | | | | | Sample Size Used :12 | | Simulations: 1000 | | | |

Table S6.
A list of models targeted by causal mediation analysis for total nitrogen in Fig. 7 and their statistical values. Abbreviations in the table mean the following: Total_N, Total nitrogen; ACME, average causal mediation effect; ADE, average direct effect; #, p <0.1; *, p <0.05.

| Total N regression models | | | | | | (I) Total_N ~ Corallococcus + Gemmatimonas + Opitutus + Sphingobium + Devosia + Candidatus.Solibacter + Bradyrhizobium + Paenibacillus |  |  |  |
|---|---|---|---|---|---|---|---|---|---|
| | | | | | | (II) Devosia ~ Candidatus.Solibacter + Sphingobium | | | |
| | | | | | | Nonparametric bootstrap Confidence Intervals (T: Sphingobium - M: Devosia) | | | |
| (I) | Estimate std. | Error | t value | Pr ( > ItI) | | Estimate 95% | CI lower 95% | CI upper | p-value |
| (Intercept) | 1.4243 | 0.3378 | 4.217 | 0.0244 | * | ACME | 0 | 0 | 0 | 1 |
| Corallococcus | 49.1327 | 14.6833 | 3.346 | 0.0442 | * | ADE | 0.79 | -0.583 | 1.72 | 0.23 |
| Gemmatimonas | 83.3747 | 28.2376 | 2.953 | 0.0599 | # | Total Effect | 0.79 | -0.583 | 1.72 | 0.23 |
| Opitutus | 60.682 | 25.6554 | 2.365 | 0.0989 | # | Prop.Mediated | 0 | 0 | 0 | 1 |
| Sphingobium | -36.6369 | 26.0637 | -1.406 | 0.2545 | | | | | |
| Devosia | -9.2243 | 16.8605 | -0.547 | 0.6224 | | Sample Size Used :12 | Simulations: 1000 | | |
| Candidatus.Solibacter | -91.0711 | 37.2804 | -2.443 | 0.0923 | # | | | | |
| Bradyrhizobium | 17.6118 | 24.0006 | 0.734 | 0.5162 | | Quasi-Bayesian Confidence Intervals (T: Sphingobium - M: Devosia) | | | |
| Paenibacillus | 51.015 | 16.7616 | 3.044 | 0.0557 | # | Estimate 95% | CI lower 95% | CI upper | p-value |
| | | | | | | ACME | 0 | 0 | 0 | 1 |
| | | | | | | ADE | 0.778 | -0.197 | 1.77 | 0.13 |
| (II) | Estimate std. | Error | t value | Pr ( > ItI) | | Total Effect | 0.778 | -0.197 | 1.77 | 0.13 |
| (Intercept) | 0.008723 | 0.003402 | 2.564 | 0.0305 | * | Prop.Mediated | 0 | 0 | 0 | 1 |
| Candidatus.Solibacter | -0.155179 | 0.387255 | -0.401 | 0.698 | | | | | |
| Sphingobium | 0.789838 | 0.502592 | 1.572 | 0.1505 | | Sample Size Used :12 | Simulations: 1000 | | |